\documentclass[twocolumn]{emulateapj}
\usepackage{natbib}
\usepackage{graphicx}
\usepackage[dvipsnames]{color}
\usepackage{float}
\usepackage{longtable}
\usepackage{enumerate}
\usepackage{graphicx}
\usepackage{soul}

\usepackage{amsmath}

\begin{document}

\title
{Scintillation based search for off-pulse radio emission from pulsars}

\author{Kumar Ravi $^{1}$, Avinash A. Deshpande$^{1}$ }
\affil{$^{1}$Raman Research Institute, Bangalore, India; raviranjan@rri.res.in; desh@rri.res.in\\}

\begin{abstract}
We propose a new method to detect off-pulse (unpulsed and/or continuous)
emission from pulsars, using the intensity modulations associated with
interstellar scintillation.
Our technique involves obtaining the dynamic spectra,
separately for on-pulse window and off-pulse region,
with time and frequency resolutions to properly sample
the intensity variations due to 
diffractive
 scintillation, and then
estimating their mutual correlation as a measure of off-pulse emission,
if any. We describe and illustrate the essential details of this
technique with the help of 
simulations, as well as real data.
We also discuss advantages of
this method over earlier approaches to detect off-pulse emission.
In particular, we point out how certain non-idealities inherent to
measurement set-ups could potentially affect estimations in earlier approaches,
 and argue that the present technique is immune to
such non-idealities. We verify both of the above situations
with relevant simulations. 
We apply this method to observation of
PSR B0329+54 at frequencies 730 and 810 MHz,
made with the Green Bank Telescope and
present upper limits for the off-pulse intensity
at the two frequencies.
We expect this technique
to pave way for extensive investigations of off-pulse
emission with the help of even existing dynamic spectral
data on pulsars and of course with more sensitive
long-duration data from new observations.
\end{abstract}


\section{Introduction}
It is the pulsed nature of the emission (as against continuous emission)
that made the discovery of pulsars (Hewish et al. 1968) possible.
Their average intensities,
if were to manifest as continuous emission,
are in most cases too weak to be detectable,
in presence of possible confusion from other continuous sources.
The pulsed emission has been studied in great detail,
and has lead to our present understanding of the physical
picture of pulsars. However, the question as to whether
pulsar radiation indeed has any intrinsic continuous component,
in addition to its distinguishing pulsed signature,
or if the periodic emission extends well beyond the main/inter-pulse windows,
have been issues of much interest since the early days of
pulsar studies.

There have been several attempts 
to detect off-pulse emission
from pulsars (as summarized in Table 3 and discussed in Section 5 
of Basu et al. 2011).
Most attempts were primarily aimed at
detection of unpulsed emission component of magnetospheric origin
(for example, Hugunein et al. 1971; Bartel et al. 1984; Perry \& Lyne 1985; 
Hankins et al. 1993; Basu et al. 2011,2012), which is
indeed the focus of 
this paper. In contrast, some 
were prompted by, and were aimed to test,
the proposition of Blandford et al. (1973) -
``existence of ghost supernova remnants around old pulsars".
Detection of unpulsed emission of magnetospheric origin is
indeed challenging, when based on apparent intensity in
the off-pulse region, particularly in presence of
a variety of unresolved astronomical sources 
and the resulting \emph{confusion}.
Such contaminants could include pulsar companions, if any,
nearby galactic/extragalactic radio sources, and diffuse
background emission, in addition to the following sources
associated with the pulsar.
They may include (e.g. as discussed by Hankins et al. 1993)
weak halos (Blandford et al. 1973),
remnants of the progenitor supernova,
shock structures or synchrotron nebulae,
and detectable bow shock. All these contaminations
are unavoidable because of finite beam-width of single-dish
telescopes and non-negligible side-lobes of interferometers.
\par
After several non-detections and some reports of
detections that were refuted subsequently,
the off-pulse emission has attracted renewed attention
with Basu et al. (2011, 2012) reporting detection of
off-pulse emission from B0525+21 and B2045-16
 based on their GMRT observations. It is worth noting,
that in their study of 20 pulsars,
including B0329+54, B0525+21 and B2045-16,
at 2.7 and 8.1 GHz using the NRAO 3-element interferometer,
Huguenin et al. (1971) found no significant unpulsed emission,
implying an upper limit of 20 mJy within 10 arcsec of the pulsar
directions. Much later, Bartel et al. (1984) made observations
of pulsars B0329+54 and B1133+16 at 2.3 GHz using Mark III VLBI,
and also ruled out continuous emission above their
detection limit (2.5 mJy). Soon after, Perry \& Lyne (1985),
reported their interferometric observations at 408 MHz,
on 25 pulsar including B0329+54 and B0525+21,
made using 76m MK 1A telescope at Jodrell Bank and the 25m telescope at Defford,
with baseline of 127 km. They claimed detection of unpulsed emission
from 4 pulsars B1541+09, B1929+10, B1604-00 and B2016+28.
However, later it became clear that B1541+09 and B1929+10
are aligned rotators (Hankins et al. 1993; Rathnasree \& Rankin 1995),
and the unpulsed emission from  B1604-00 and B2016+28 were
shown to be from unrelated background sources
(Strom \& Van Someren Greve 1990; Hankins et al. 1993).

The recently reported detections of off-pulse emission
(Basu et al. 2011; 2012) from two long period pulsars
B0525+21 (3.75 s) and B2045-16 (1.96 s) are based on
the imaging mode of GMRT, and also at two frequencies (325 and 610 MHz).
Although the authors have discussed some effects
that could potentially contaminate off-pulse region
with leakage from the emission that is otherwise
confined to the main-pulse window, and have attempted
some tests based on which they claim absence of such leakage.
We consider these tests inadequate to rule out ``leakage",
since there are a few different aspects, associated with
commonly employed receiver setups, that have noticeable
potential for undesirably spilling the main pulse contribution
across off-pulse region.

Ideally, we need a method that is
immune to such contamination, as far as possible,
 while making reliable estimation
of possible off-pulse or unpulsed emission intrinsic to pulsar.

Owing to their compact size and pulsed emission, pulsars have
been an excellent probe of the ISM since their discovery.
Primarily, they have revealed the distribution of free electron
in the Galaxy, through direct measures of column density
(from the observed dispersion) and spatial distribution of
electron density irregularities (from scintillations and
angular/temporal broadening as a result of scattering).
The highly polarized nature of their radiation also
allows Faraday Rotation measurements,
sampling the magneto-ionic component of the intervening medium,
and their pulsed nature facilitates some of the clearest
measurements of HI absorption along their sight-lines.
Of these, the diffractive scintillation effects
are readily observable in pulsar directions thanks
to their tiny angular sizes, and become apparent only 
in the cases of some extra-galactic sources having the
required compact angular size, such as in early phases of
$\gamma$-ray burst (GRB) afterglow sources 
(see for example, Frail et al. 1997; Macquart \& de Bruyn 2006).

The diffraction induced chromatic
modulation of intensity, when combined with relative motions,
translates to intensity variations across time and frequency.
Similarly it is only the pulsed nature of the radio pulsars which
makes the dispersion effect measurable,
and also to reveal the temporal broadening due to scattering. However,
the latter can be probed indirectly
via other manifestation of scattering (such as decorrelation scales
in frequency and/or angular broadening), even in the case of
continuous sources. The {\it camaraderie} between the pulsars and
the interstellar medium is indeed reciprocal. For example,
ultra-high angular resolution probe of pulsar emission
is made possible by the ISM acting as a lens.
This was first pointed out by Lovelace (1970) and has been
followed-up by many (e.g. Cordes \& Wolszczan 1988; Pen et al. 2014;
and references therein). Here,
the diffractive/refractive effects due to large scale irregularities
are considered as providing interstellar interferometric measurements
capable of resolving even magnetospheric emission regions of pulsars.
The refractive effects leading to multiple imaging
manifest themselves as fine-scale corrugations
or drift patterns within scintles in the dynamic spectra resulting
from diffractive scintillations (e.g. Wolszczan and Cordes 1987,
Gupta et al. 1994,1999).

In this paper, we present a technique which advantageously uses
such interstellar-scale telescope for search and detection of
unpulsed emission, if any, from pulsars.
Our technique (described in Section 2) is based on diffractive 
interstellar scintillation (DISS) 
and its correlated imprint on the pulse 
intensity and any off-pulse emission
intrinsic to the pulsar, and  
has the potential
for providing more reliable measurement of intrinsic 
off-pulse/unpulsed emission,
without needing conventional interferometric measurements,
i.e. which are possible even with single-dish observations.

In Section 3,
we demonstrate sensitivity of our technique using 
simulated dynamic spectra over wide
band, and assess its
immunity to various known sources of
contamination in the off-pulse region. 
Discussion of one such potential contaminant is given
in Appendix A. 
The details of the DISS simulation
are presented in the Appendix B.
In Section 4, 
we illustrate application of our technique to real data, using
observations on B0329+54 at two radio frequencies. 
We summarize the main conclusions of our paper in Section 5.


\section{Scintillation-based technique for search/detection of unpulsed emission from pulsars}

In this section, we present a new technique based on diffractive
interstellar scintillation (DISS) and assessment of correlation
between dynamic spectra for the pulse and
off-pulse intensities.
It effectively renders measurements with fine angular resolution
offered by interstellar diffraction
to distinguish pulsar emission region from sources of confusion,
even in close proximity to the pulsar. 
This DISS correlation criterion effectively
and readily discriminates against all discrete and 
diffuse radio emissions on angular scales
larger than that of the pulsar magnetosphere, since they will
be devoid of DISS imprint in their dynamic spectra, 
let alone show any correlation with pulse intensity variations.
Any confusing compact source,
unresolved by the observing telescopes,
and compact enough to show DISS,
will show a dynamic spectral signature,
i.e., the scintillation pattern, significantly different from
that associated with the pulsar emission. In fact,
differences between the scintillation patterns associated with
even the different components within the pulse profile have
been probed to assess spatial separation,
if any, between the apparent sites of emission (Cordes et al. 1983).

If indeed, a pulsar has a component of intrinsic
emission that is unpulsed/continuous, we expect its
intensity modulation due to interstellar scintillation
to be closely related to, if not matching, that of
the pulsed component. 
For the desired correlation to exist between the
diffractive scintillation spectra of intensities in the
two longitude regions, the spatial transverse separation between
the associated emission regions should ideally be well
within the equivalent spatial resolution
of the interstellar aperture/interferometer at work.
As Cordes et al. (1983) have already noted, the spatial
scale $S_\mathrm{d}$ of the diffraction pattern in the observer plane
also (reciprocally) defines the associated spatial
resolution at the source distance.

A suitable data set for implementation of our technique
is, in general, an appropriately sampled data cube of intensity $I(\nu,t,\phi)$,
as a function of rotational
longitude $\phi$, radio frequency $\nu$ and time $t$,
and over wide frequency and time spans of, say, $\Delta\nu_\mathrm{BW}$ 
and $\Delta t_\mathrm{obs}$, respectively.

The two dynamic spectra, $I_\mathrm{on}(\nu,t)$ and $I_\mathrm{off}(\nu,t)$,
to be tested for mutual correlation,
are to be constructed for the apparent average intensity
across (i) an appropriate number of bins spanning or within
the pulse window, and (ii) a chosen set of bins or longitude range
in the off-pulse region which is well-separated from the pulse window.
All of the (dynamic) spectra here are assumed to be already 
corrected for any non-uniformity in spectral response of the observing 
system within the observed band.\footnote{An estimate of the required 
{\it normalized} spectral 
gain response ($G(\nu)$), to be used for dividing all the observed spectra, 
can be made by averaging the observed off-pulse spectra over the entire time span
of observation to first obtain a mean uncalibrated spectrum 
$<S_\mathrm{off}>(\nu)$,
and then normalizing it with band-averaged intensity $\bar{S}_\mathrm{off}$, 
such that $G(\nu) = <S_\mathrm{off}>(\nu)/{\bar{S}_\mathrm{off}}$.}

Sensitivity in the estimation of correlation depends
on the signal-to-noise ratio in estimation of
the two dynamic spectra, and the degrees of freedom
provided by the richness in the dynamic spectra, quantifiable
to the first order in terms of number of scintles.
Naturally, scintillation dynamic spectra obtained from
longer duration observations with wide spectral coverage
are desired, if not essential. The dynamic spectral
resolutions in time and frequency, say, $\delta t_\mathrm{res}$
and $\delta\nu_\mathrm{res}$, respectively,  need to be adequately
finer than the respective decorrelation scales ($t_\mathrm{s}$ and $\nu_\mathrm{d}$),
which together characterize the average {\it size} of scintles.
The dynamic spectra, therefore, are to be smoothed
optimally to reduce the uncertainty in estimation of
the intensity variations due to scintillation, without
washing out details in the ISM induced diffractive
variation of interest.

In practice, the dynamic spectra are not free of
(additive) random noise in estimating intensity at each
pixel in the time-frequency plane, but the magnitude of
this noise is expected to be
largely consistent with the system temperature
and the integration employed (quantified by the
relevant time-bandwidth product).
Thus, in general, $I_\mathrm{on}(\nu,t) = I^p_\mathrm{on}(\nu,t) + U_\mathrm{on}(\nu,t)$,
and $I_\mathrm{off}(\nu,t) = I^p_\mathrm{off}(\nu,t) + U_\mathrm{off}(\nu,t)$,
where the $U_\mathrm{on}$ and $U_\mathrm{off}$ represent random noise
(with zero mean and standard deviations $\sigma_\mathrm{on}$ and $\sigma_\mathrm{off}$
respectively),
which is uncorrelated from pixel to pixel, and contaminates
the respective underlying pulsar dynamic spectra $I^p_\mathrm{on}$ and
$I^p_\mathrm{off}$.  These delta-correlated noise contributions,
of magnitude $\sigma^2_\mathrm{on}$ and $\sigma^2_\mathrm{off}$,
will be clearly noticeable as such at zero-lag in
the respective auto-correlation functions, $ACF_\mathrm{on}$
and $ACF_\mathrm{off}$, of the dynamic spectra, on top of the
the otherwise smoothly varying auto-correlations
of $I^p_\mathrm{on}$ and $I^p_\mathrm{off}$, respectively.
Hence, the zero-lag auto-correlation of the underlying
intensity variation is estimated routinely by interpolation
from correlations at adjacent lags.

The average cross-correlation between the intensity variations
in two dynamic spectra, $I_\mathrm{on}(\nu,t)$ and
$I_\mathrm{off}(\nu,t)$, defined as
\begin{equation}
CC(0,0) = <\delta I_\mathrm{on}(\nu,t)\,\,\,\delta I_\mathrm{off}(\nu,t)>
\end{equation}
at zero lags, is to be assessed for significance against uncertainties,
where $\delta I_\mathrm{yy}(\nu,t) = I_\mathrm{yy}(\nu,t)-<I_\mathrm{yy}>$
for {\it yy} state ({\it on} or {\it off}),
and $<x>$ indicates ensemble average of $x$ across the span
of ($\nu,t$).
The uncertainty in the estimated correlation, in the best case
(i.e. dynamic spectra free of noise and other undue contaminants),
will be dominated finally by the finiteness of available scintle statistics.
In case of detection of significant correlation, the off-pulse
emission intensity as fraction $\eta$ of on-pulse intensity
can be estimated as
\begin{eqnarray}\nonumber
\eta &=& \frac{<\delta I_\mathrm{on}(\nu,t)\,\,\,\delta I_\mathrm{off}(\nu,t)>}
            {<(\delta I_\mathrm{on}(\nu,t))^2>} \\
     &=& \frac{CC(0,0)}{AC_\mathrm{on}(0,0) - {\sigma}^2_\mathrm{on}}
\end{eqnarray}
where $AC_\mathrm{on}(0,0)$ is the average zero-lag auto-correlation
of (on-)pulse intensity variations, which includes
the variance $\sigma^2_\mathrm{on}$ of the delta-correlated noise
$U_\mathrm{on}(\nu,t)$.

In the discussion so far, the apparent intensity
fluctuations across the dynamic spectrum for the on-pulse
region are, in an ideal case, assumed to be primarily
a manifestation of
the interstellar scintillations across the observing band.
However, a finite but small part of these
may be due to 1) variations in the system noise, including the
sky noise (other than that from the pulsar), in addition
to 2) the contribution from aliased spectral range, if any.
The former additive contributions equally affect
the dynamic spectrum for the off-pulse region, and
may undesirably contribute
to the apparent correlation between the dynamic spectra.
It is therefore important that the {\it on-pulse} dynamic spectrum
$I_\mathrm{on}(\nu,t)$ is obtained after subtraction of $I_\mathrm{off} (\nu,t)$
from the corresponding spectrum for the on-pulse region.
The version of $I_\mathrm{off}(\nu,t)$, to be used for subtraction here,
should be for intensity averaged over the entire off-pulse region,
as far as possible. In case of any genuine unpulsed intensity with
correlated variations with those for on-pulse region,
the suggested subtraction would result in an under-estimation of
$\eta$ by an amount ${\eta}^2$. On the other hand, even
intrinsic variations in the off-pulse region that are uncorrelated
between the two dynamic spectra would be unduly subtracted
from the on-pulse dynamic spectrum, and would introduce
a negative bias in $\eta$ estimate. The magnitude of such bias
is given by the ratio of variance ${\sigma}^2_U$
of these uncorrelated variations in the off-pulse spectrum
to that for variations in the on-pulse spectrum
(i.e. ${\sigma}^2_U/({AC_\mathrm{on}(0,0) - {\sigma}^2_\mathrm{on}})$).
In any case, the negative bias will be limited to ${\eta}^2_\mathrm{max}$,
where $\eta_\mathrm{max}$ is as defined later in the Equation 6.
The advantage of thus removing any common unpulsed intensity
variations, either due to sky or system,
from $I_\mathrm{on}(\nu,t)$, in terms of obtaining a more reliable
estimate of $\eta$, overwhelms the undesirability of the
the mentioned bias, which is expected to be insignificant any way.

Of course,
any intrinsic variability in the pulsar intensity would 
leave an unavoidable (multiplicative) imprint in the dynamic spectrum. 
The spectral scales of intrinsic variability  
are expected to be much wider than those associated with 
interstellar scintillation. There is no a priory basis yet
for expecting the possible unpulsed component, if any, 
to have correlated intrinsic variability. Hence, in general,
any independent intrinsic variability of intensities in the two
regions would reduce the net cross-correlation, and in any case,
increase the uncertainty in the estimation of the unpulsed
intensity.
Fortunately, any pulse-to-pulse variations in intrinsic intensity are 
expected to average out, with suitable temporal smoothing 
of the dynamic spectrum ($P\ll\Delta t_\mathrm{res}< t_\mathrm{s}$). 
Any residual variation, on time scales shorter than $\Delta t_\mathrm{res}$,
would be indistinguishable from the random uncertainty in estimation
of the dynamic spectral elements. The combined magnitude of
these fluctuations would be readily apparent in the auto-correlation
function across the first few time-lags, as the delta-correlated
contribution.
In comparison, the auto-correlation due to scintillation-induced
intensity variations is expected to decorrelate on a relatively
longer time-scales ($t_\mathrm{s}$).

\par
The expected implicit linear inter-relationship between
the patterns (after removing the respective mean values), 
assessed through formal cross-correlation,
can be modeled explicitly as follows
\begin{equation}
\delta I_\mathrm{off}(\nu,t) = \eta\,\delta I_\mathrm{on}(\nu,t) + U(\nu,t)
\end{equation}
where, the first term on the right-side is the best-fit model, and
$U(\nu,t)$ is 
the apparent deviation or 
the part of
observed off-pulse dynamic spectrum that is
uncorrelated in time and frequency with $\delta I_\mathrm{on}(\nu,t)$,
with its nominal mean $\left<U(\nu,t)\right>=0$,
and other quantities as defined earlier.
The uncorrelated part $U(\nu,t)$ includes also
any measurement uncertainties in {$I_\mathrm{off}$ and also
the model $\eta\, I_\mathrm{on}$.}
In the above formulation, as in the Equation 2,
$\eta$ is a measure of the ratio $\delta I_\mathrm{off}/\delta I_\mathrm{on}$.

The uncertainty $\sigma_\mathrm{\eta}$ in its estimate
can be expressed as
\begin{equation}
\sigma_\mathrm{\eta}=\frac{\sigma_\mathrm{<U>}}{\sqrt{<(\delta I_\mathrm{on}(\nu,t))^2>}}
\end{equation}
where $\sigma_\mathrm{<U>}$ is the reduced
uncertainty in the mean of $U(\nu,t)$,
and is related to standard deviation in $U(\nu,t)$ as
\begin{equation}
\sigma_\mathrm{<U>} = {\sigma}_U \sqrt{\frac{1}{N_\mathrm{eff}}} 
\end{equation}
where $N_\mathrm{eff}$ is the effective size of the ensemble.
The $<U>$ and ${\sigma}_U$ are, in practice,
computed using all of the $N$ samples available in
the dynamic spectral array, including $U(\nu,t)$.
The total number of points $N$ in these arrays
is equal to $N_\mathrm{\nu0}N_\mathrm{t0}$,
where $N_\mathrm{\nu 0}$ is the number of spectral channels and
 $N_\mathrm{t0}$ number of time bins/sections in the dynamic spectrum.
However, since all the points/pixels in the dynamic spectrum
are not independent, particularly when
the random measurement noise is much smaller than
the intensity variations due to scintillation.
Hence, in such cases, $N_\mathrm{eff}$ is often much smaller than $N$,
and represents rather the number of independent
samples in the dynamic spectrum.  We have used the number
of scintles as defining $N_\mathrm{eff}$, so that our uncertainty
estimate $\sigma_\mathrm{\eta}$ corresponds to worst-case error.
The definition of number
of scintles, as given in Cordes \& Lazio (1991), is
$N_\mathrm{eff}=N_\mathrm{t}\times N_\nu$, where
$N_\mathrm{t}=1+\kappa ({\Delta t_\mathrm{obs}}/{t_\mathrm{s}})$,
$N_\nu=1+\kappa ({\Delta\nu_\mathrm{BW}}/{\nu_\mathrm{d}})$,
where $\kappa$ is an empirically obtained number
(we can call it {\it filling factor}), lies in range $0.1-0.5$.
If $N_\mathrm{eff}$ for one spectrum is different from that for the other,
we use the geometric mean of the two $N_\mathrm{eff}$ values.

For dynamic spectra spanning long durations, explicit
attention would be needed to examine if they are affected
by possible slow variations in
pulse intensity within the span, due to intrinsic variations
and/or originating from extrinsic reasons, including
refractive scintillations and any instrumental gain variations
that remain to be corrected.
The correlation scales across frequency for these are
expected to be generally wide. Hence, any contamination
in the off-pulse region, as mentioned above,
is likely to be modulated the same way,
resulting in spurious correlation corrupting the correlation of interest.
It may become necessary therefore to either estimate
slow modulation, and correct at least the on-pulse
dynamic spectra accordingly, or estimating the correlation
or $\eta$ using dynamic spectra of shorter spans at a time,
repeating the analysis for each of such sections separately,
and then computing a weighted average of $\eta$,
combining independent estimates made using subsets of data.

Before proceeding further, we wish to draw attention
to a particular ready utility of the dynamic spectra
of the apparent intensity variations in the on-pulse
and off-pulse regions.
We argue that, regardless of the details of contamination, and
the presence or the lack of correlation between
the two dynamic spectra, it is possible to define a hard
upper-limit for the unpulsed intensity, as
\begin{eqnarray}\nonumber
\eta_\mathrm{max} &=& \sqrt {\frac{<\delta I_\mathrm{off}(\nu,t)^2>}
            {<(\delta I_\mathrm{on}(\nu,t))^2>} }\\
         &=& \sqrt{\frac{AC_\mathrm{off}(0,0) - \sigma^2_\mathrm{off}}
            {AC_\mathrm{on}(0,0) - \sigma^2_\mathrm{on}}}
\end{eqnarray}
where $AC_\mathrm{off}(0,0)$ is the average zero-lag auto-correlation
of observed intensity variations in the off-pulse region, which includes
the variance $\sigma^2_\mathrm{off}$ of the delta-correlated noise
$U_\mathrm{off}(\nu,t)$. 
When $AC_\mathrm{off}(0,0) \gg \sigma^2_\mathrm{off}$,
fractional uncertainty ($1\sigma$) in $\eta_\mathrm{max}$ would be $1/\sqrt{2N_\mathrm{eff}}$. 
However, even when $\delta I_\mathrm{off}(\nu,t)$ appears to consist of 
only delta-correlated noise, i.e. $\eta_\mathrm{max} \approx 0$, the uncertainly
would at best be limited to $\sigma_\mathrm{off}/\sqrt{N<(\delta I_\mathrm{on}(\nu,t))^2>}$.

\subsection{Implications of relative location of possible off-pulse emission region}

In general, the apparent emission in the off-pulse region would be
a combination of the intrinsic and confusing sources of continuous
emission, and the discussed correlation would be correspondingly partial,
but providing a measure of the intrinsic component (spatially confined
within the transverse scale $S_\mathrm{d}$).

Given the form of spatial distribution
of electron density irregularities in the ISM 
as detailed in Appendix B,
this spatial resolution scale $S_\mathrm{d}$, 
same as
the diffraction pattern scale, is given by the following
relation (Armstrong et al. 1995). 
\begin{equation}
S_\mathrm{d}=\left[8\pi r_\mathrm{e}^2\lambda^2C_\mathrm{N}^2zf(\alpha)/(\alpha+1)\right]^{-1/\alpha}
\end{equation}
where, $r_\mathrm{e}$ is the classical radius of electron,
$\lambda$ is the radio wavelength,
$\alpha$ = $\beta - 2$,
and  $z$ is the effective propagation distance through 
the ISM.\footnote{
For uniformly distributed scattering, $z$ would
correspond to the distance to the pulsar.
For Kolmogorov turbulence, $\beta=11/3$
($\alpha=5/3$) and the numerical value of the
function $f(\alpha)$ is $\approx1.12$.}

The ISM parameters in the above equation are not directly
measurable, although can be 
estimated.\footnote{The diffractive scintillation time-scale $t_\mathrm{s}$ (decorrelation time)
is directly related to $S_\mathrm{d}$, where $t_\mathrm{s} = S_\mathrm{d}/V$,
but the equivalent velocity $V$ of the medium relative
to the pulsar sight-line is not independently
known in most cases.
On the other hand, the associated angular scatter
broadening $\theta$, which ultimately limits the
resolution in imaging observations,
also relates to the above spatial scale, as
$\theta = \lambda/2\pi S_\mathrm{d}$.}
A related and more readily measurable quantity is
the scintillation decorrelation bandwidth $\nu_\mathrm{d}$, or
alternatively the temporal scatter broadening $\tau$
of the pulse, where
$\tau = z{\theta}^2/2c \approx 1/2\pi\nu_\mathrm{d}$, and $c$ is the
speed of light.
Thus, $S_\mathrm{d}$
can be estimated from $\nu_\mathrm{d}$ measurement,
as\footnote{The form of this expression is consistent
with that in Equation 13 of Cordes et al. (1983).}

\begin{equation}
S_\mathrm{d} = \sqrt{z{\lambda}^2 \nu_\mathrm{d}/4c\pi} = r_\mathrm{f} \sqrt{\nu_\mathrm{d}/2\nu}
\end{equation}
where $r_\mathrm{f}$(=$\sqrt{z\lambda/2\pi}$) is the Fresnel scale,
and $\nu$ is the observing radio frequency.

A positive result in our proposed correlation test 
would not only conclusively confirm
the claimed detections, but would constrain the apparent size and
spatial separation, at the so-called ``retarded emission time"
(Cordes et al. 1983),
between the corresponding emission regions, and the level
of correlation would provide clues on the relative spatial
separation. A negative result, on the contrary, would not
necessarily imply absence of unpulsed emission, unless
the resolution scale $S_\mathrm{d}$ is large enough to cover
the entire spatial extent within which emission can be
considered as intrinsic to the pulsar.
Considering the maximum
separation between relevant emission regions to be
the so-called light-cylinder radius $r_\mathrm{L}$ ($=cP/2\pi$),
the above requirement implies that $r_\mathrm{L}\lesssim S_\mathrm{d}$,
where $P$ is the pulsar period\footnote{
For the spin periods in the range 1.4 ms$-$11.8 s,
the range of light-cylinder radius corresponds to
$\sim10^4-10^8$m.}.

This condition can also be expressed as 
$cP/2\pi \lesssim r_\mathrm{f} \sqrt{\nu_\mathrm{d}/2\nu}$,
assessment of which would require an estimate of $z$,
in addition to that of the decorrelation bandwidth $\nu_\mathrm{d}$.
Although independent distance measurement is
desired, $z$ estimated from dispersion measure would
also render useful for the present purpose.
It is worth emphasizing that the above stated condition
is not model dependent, i.e. independent of $\beta$.

For a given pulsar,
i.e. given $P$, $z$ and $C_\mathrm{N}^2$,
the observing frequency $\nu$ can be chosen suitably,
to see if the above condition can be met.
The condition is more likely to be satisfied in cases of
higher frequency probe of scintillation patterns for
relatively nearby short-period pulsars.\footnote{
The underlying basic dependencies,
as in the Equation 7, imply
that the spatial resolution scale $S_\mathrm{d}$
broadens with increasing radio frequency ($\nu$) and
with decreasing integrated scattering measure $C_\mathrm{N}^2z$.
This diffraction pattern scale, in the weak scattering regime at
adequately high frequencies, would of course saturate to its
upper limit, that is the Fresnel scale $r_\mathrm{f}$,
equal then to the refractive scale at its lower limit.
}

In any case, if any intrinsic unpulsed emission were to originate
within the angular scale $S_\mathrm{d}/z$ around the pulsar, we expect
to find the expected correlation signature.
Although such
cross-correlation (at {\it zero-lag}) is expected to fall significantly and
rapidly, as $\exp[ - (\Delta S/S_\mathrm{d})^{(\beta -2)} ]$ (Cordes et al. 1983),
with increasing separation $\Delta S$. However, if the separation, even if large
(i.e. many times $S_\mathrm{d}$), happens luckily to be near parallel
(within angle $(S_\mathrm{d}/\Delta S)^c$, for $S_\mathrm{d} \leq \Delta S$), then again
significant cross-correlation would be expected, but now at time-lag
$\Delta t \sim t_\mathrm{s} \Delta S/S_\mathrm{d})$, if the scattering {\it transfer function}
can be considered as essentially {\it frozen} over those time scales.
The above considerations necessitate exploration of the discussed 
correlation over a range of lags in both time and frequency, as we
do in our tests and analysis to follow.

\section{Tests with Simulated Dynamic Spectra: Assessment of sensitivity and immunity}

Here, we illustrate application of our technique to simulated
scintillation dynamic spectra, and assess its performance, in terms of
ability to reliably estimate off-pulse/unpulsed intensity 
intrinsic to the pulsar, 
and immunity to potential contaminants in the off-pulse region.

\begin{figure*}
\centering
\includegraphics[scale=0.35]{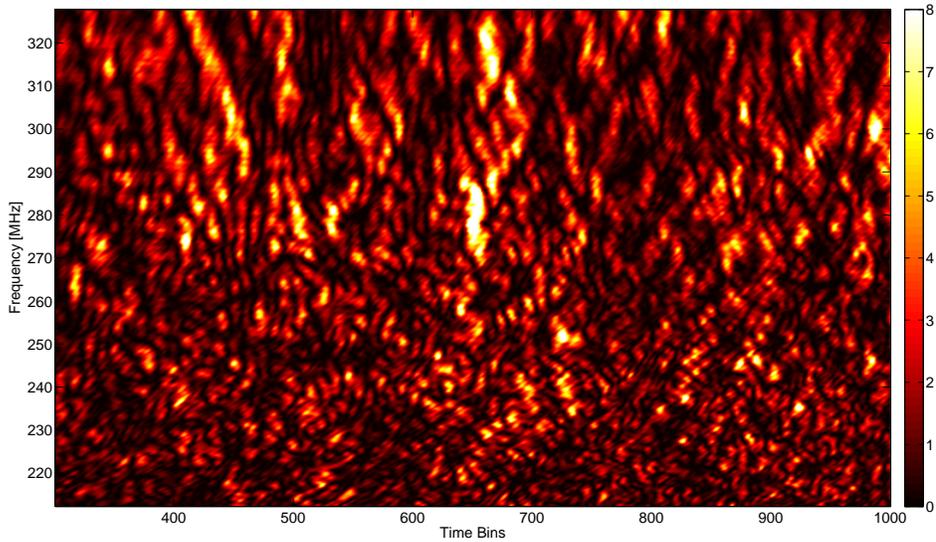}
\caption{The simulated dynamic spectrum across
a wide spectral span (bandwidth $7\times16.5$ MHz)
around a central frequency of 270 MHz,
but $C_\mathrm{N}^2$, $D$ and $S_\mathrm{d}$ chosen such that this
dynamic spectra have $\nu_\mathrm{d}$ and $t_\mathrm{d}$
similar to that corresponding to central frequency 810 MHz
for PSR B0329+54. The portion shown here is about one fourth
of the simulated time span.}
\label{simul_on_big}
\end{figure*}

\begin{figure*}
\centering
\includegraphics[scale=0.30]{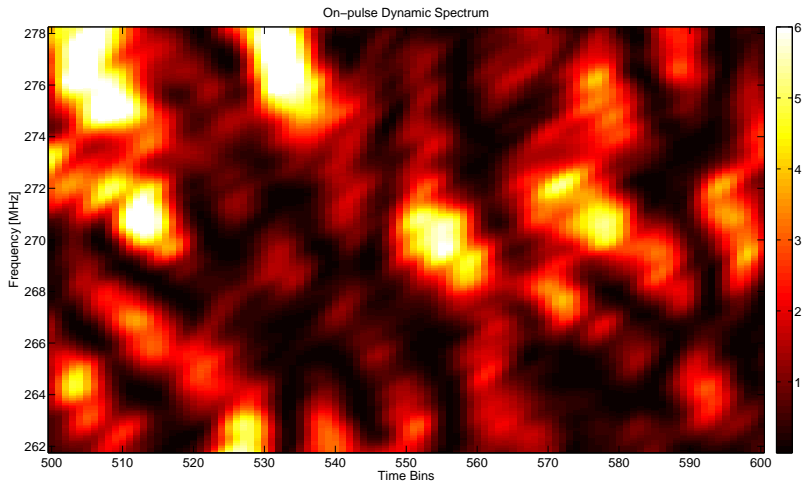}
\caption{The simulated on-pulse dynamic spectrum
 across the observing band,
 obtained from scintillation pattern shown Figure 1.}
\label{simul_on_zoom}
\end{figure*}

As mentioned in Section 2, and illustrated in Appendix A,
one of the subtle contamination of the off-pulse region could
come from genuine main-pulse signal itself, if it is not
adequately filtered out from the spectral regions
beyond the observing band. 
These aliased contributions (from possible image bands relevant to
heterodyning, and regions inadequately attenuated by band-defining filter
before digitization) appear at
longitudes that are, in general, offset from the main-pulse
window (see Figure A1), 
depending on dispersion measure and frequency separation.

Fortunately, the scintillation-induced intensity pattern
would significantly differ for spectral separations
larger than the decorrelation scales, $\nu_\mathrm{d}$, particularly when
$\nu_\mathrm{d}\ll\Delta\nu_\mathrm{BW}$,
and
even the overall shape
and sizes of the scintles (characterized by
the decorrelation scales, $\nu_\mathrm{d}$ and $t_\mathrm{s}$),
themselves vary systematically with $\nu$,
more rapidly with decreasing frequency.
Any aliased contribution from other bands will have their own
different scintillation-induced imprint, and hence,
is not expected to contribute to any significant net correlation.
This forms the basis of our expectation for potential immunity
of our scintillation correlation method against
aliasing-induced contribution which disguises as off-pulse emission,
and we assess it by using simulated dynamic spectrum over a
spectral range several (7) times the nominal bandwidth of observation.

\begin{figure*}
\centering
\includegraphics[scale=0.30]{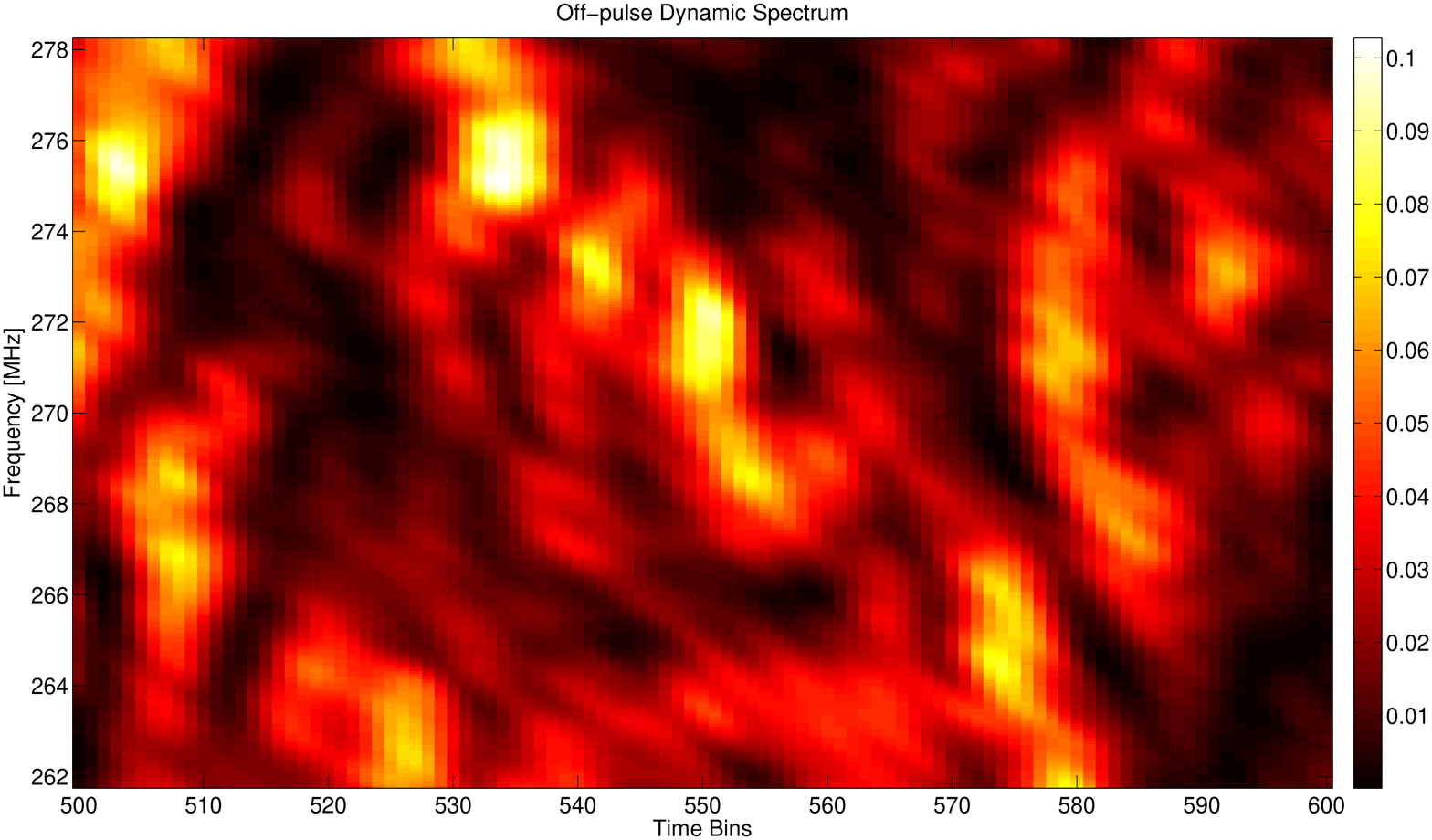}
\caption{The simulated {\it off-pulse} dynamic spectra,
corresponding to the {\it observing band} as in Figure 2, 
but obtained by
including only
the aliased contribution from the two adjacent bands around the
observing band (taken from the simulated wide-band dynamic
spectrum for the main pulse as shown in Figure 1).
This corresponds the case (ii), i.e. no genuine off-pulse/unpulsed
emission component, but contains only the contributions
due to aliasing, disguising as apparent variations
in the unpulsed intensity in the off-pulse region. }
\label{simul_off_zoom}
\end{figure*}

\begin{figure}
  \centering
  \begin{tabular}[b]{@{}p{0.44\textwidth}@{}}
    \centering\includegraphics[width=1\linewidth]{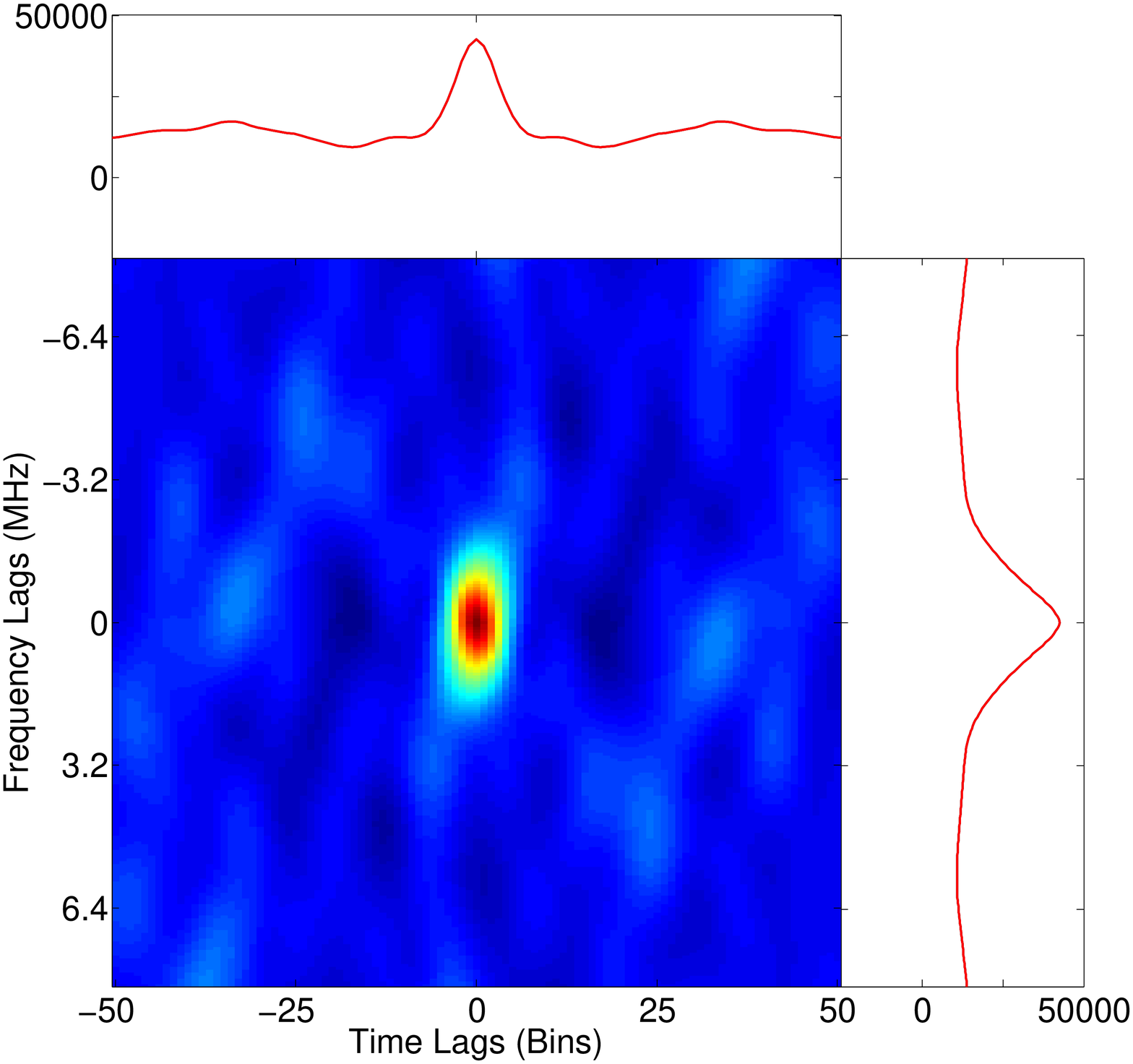} \\
    \centering\includegraphics[width=1\linewidth]{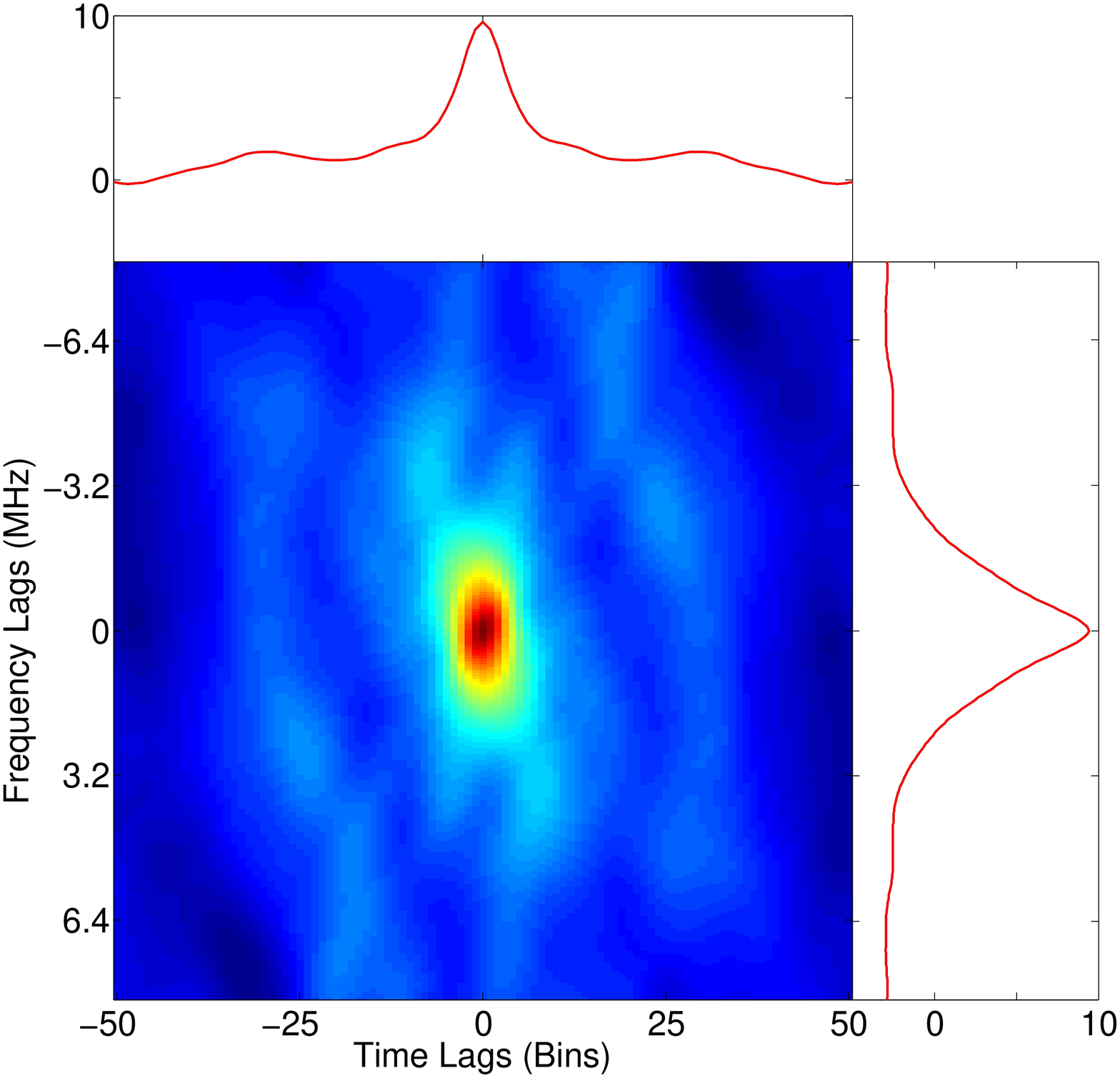} \\
  \centering\includegraphics[width=1\linewidth]{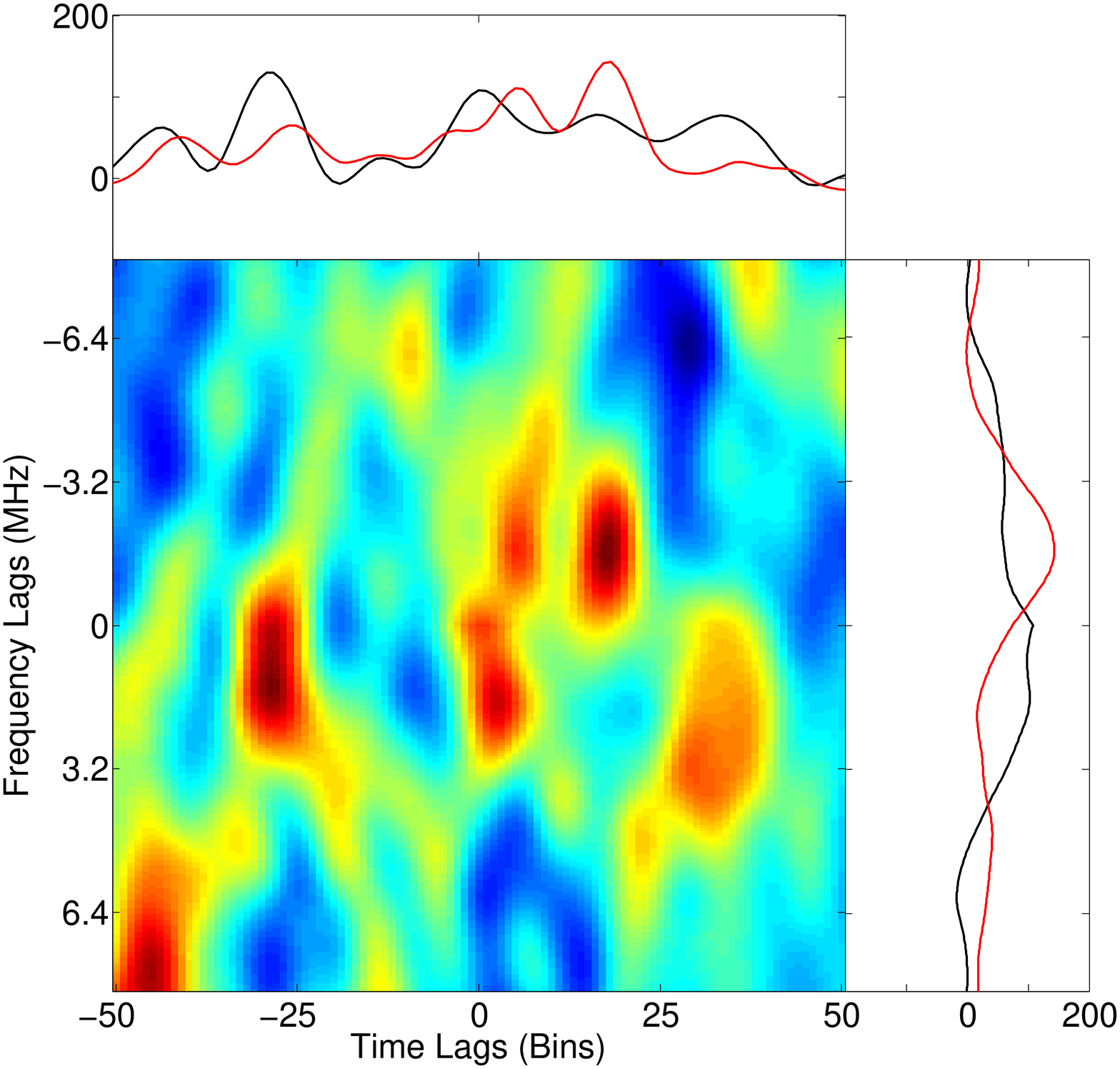} \\
  \end{tabular}
  \caption{\emph{Top:} Auto-correlation map of simulated on-pulse dynamic spectra of Figure 2.
  \emph{Middle:} Auto-correlation map of simulated off-pulse dynamic of Figure 3.
  \emph{Bottom:} Cross-correlation map of on-pulse and off-pulse dynamic spectra.
  In all these three sub-figures, the top plots are 1-D cut at zero time lag (black)
  and 1-D cut passing through maximum of the 2-D correlations(red) and side plots
  are that corresponding to frequency lags. The frequency-lag and the time-lag are
  in units of 64.45 kHz and $\sim$ $t_\mathrm{s}/5$ seconds, respectively. 
  In 2-D color-maps, color \emph{dark red} corresponds to highest magnitude and \emph{dark blue} to lowest magnitude.}
  \label{corr_simul_no_shift}
\end{figure}
\begin{figure}
  \centering
      \begin{tabular}[b]{@{}p{0.45\textwidth}@{}}
    \centering\includegraphics[width=1\linewidth]{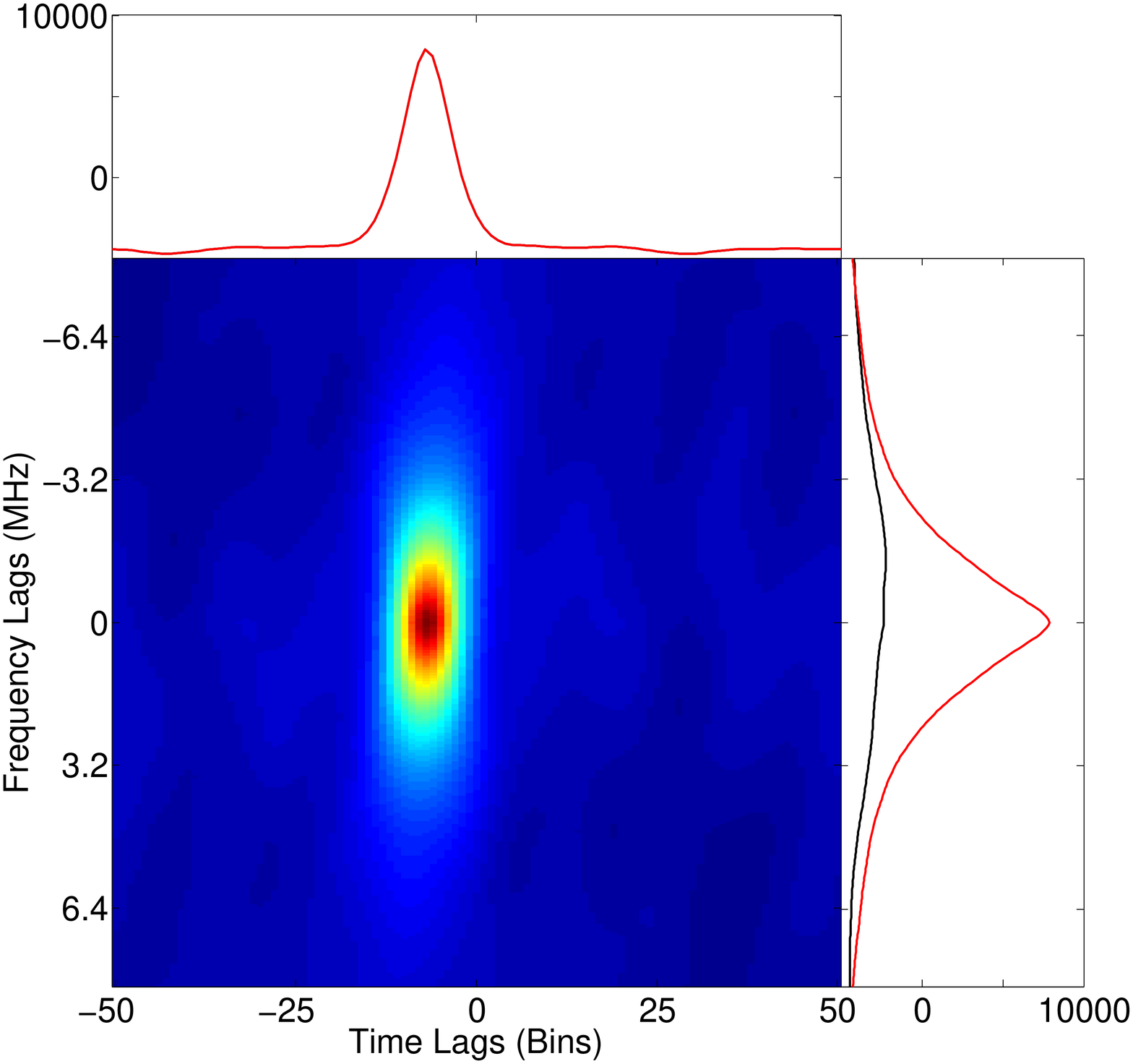} \\
    \centering\includegraphics[width=1\linewidth]{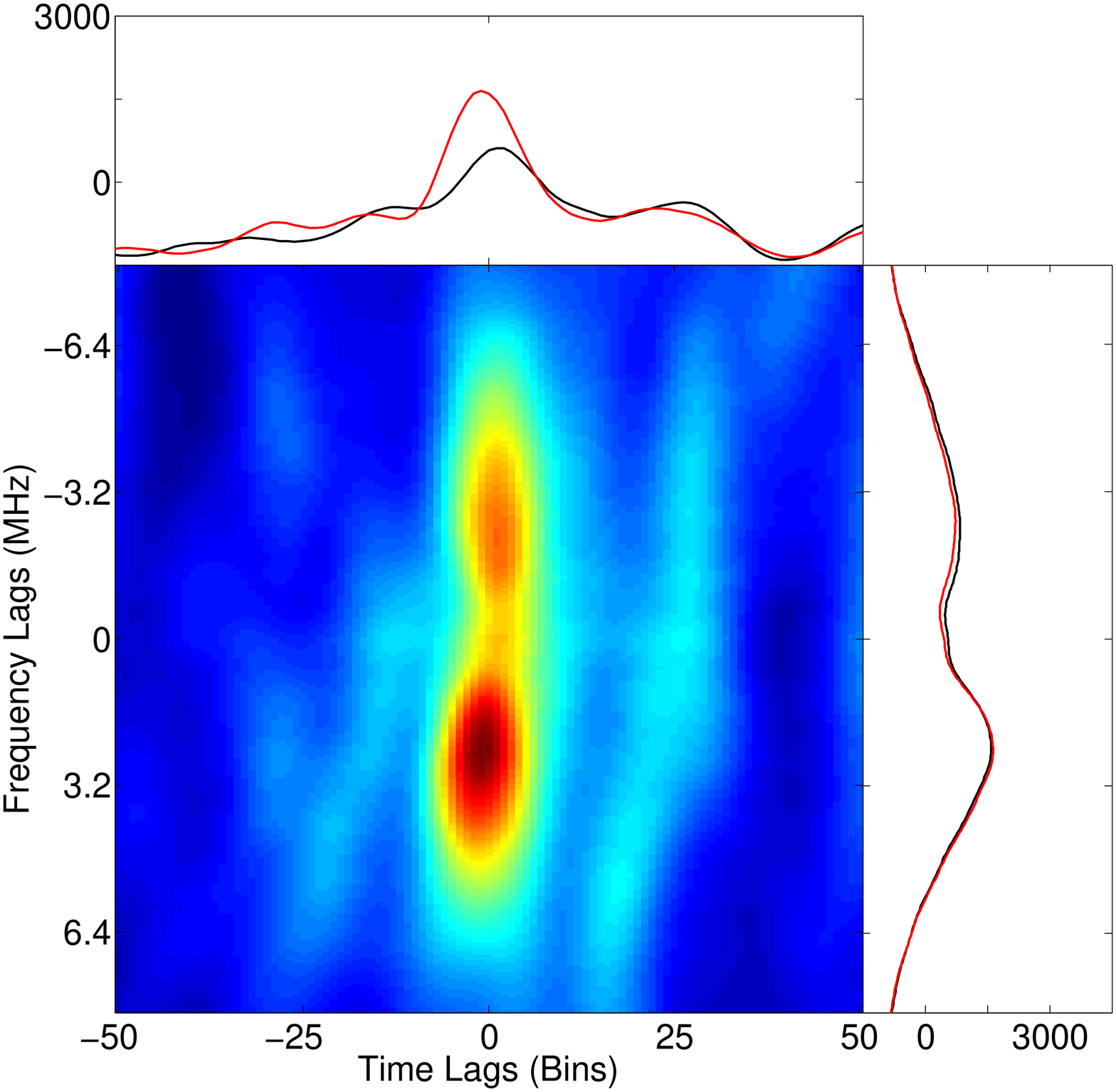} \\
    \centering\includegraphics[width=1\linewidth]{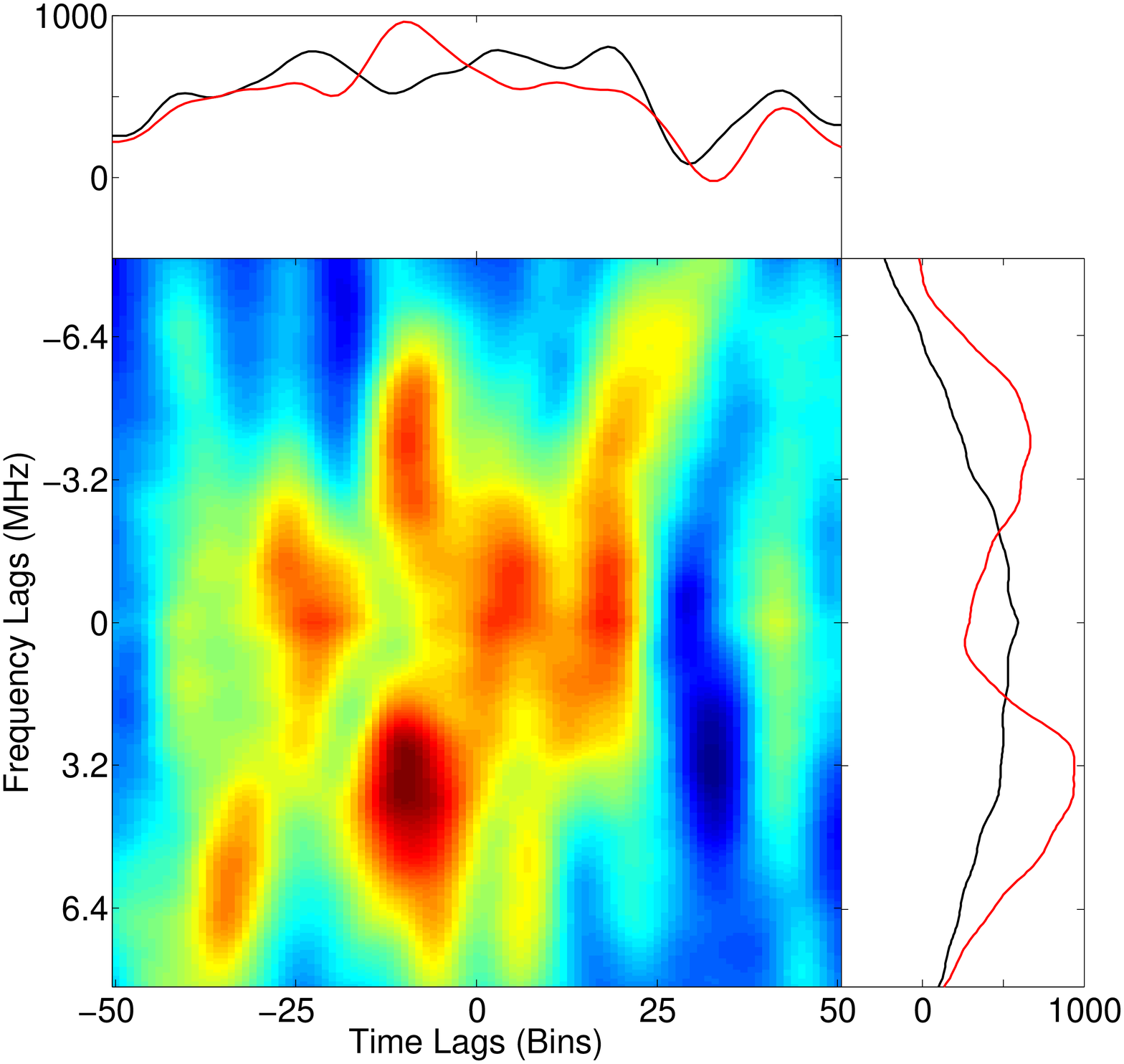} \\
  \end{tabular}
\caption{Cross-correlations of simulated on-pulse dynamic spectra and various off-pulse dynamic spectra. The off-pulse dynamic spectra are made from assuming their source separated from that of on-pulse by $0.1R_{LC}$ in, \emph{top:} x-direction (our
assumed direction of pulsars motion), \emph{middle:} y-direction and \emph{bottom:} at $45^\circ$ to x-direction. All other notions in these figures are same as those mentioned in the caption of Figure 4.}
\label{corr_simul_shift}
\end{figure}

A detailed description of our simulation of diffractive
scintillation is presented in Appendix B,
and resultant dynamic spectrum  
spans 115.5 MHz (7x16.5 MHz) centered at 270 MHz, and
covers a duration of about $1000 t_\mathrm{s}$ seconds.
The time and spectral sampling here is $\sim t_\mathrm{s}/4$ seconds and
64.45 kHz ($\sim \nu_\mathrm{d} /20$), respectively.


This simulated dynamic spectrum,
is treated as directly corresponding to
an {\it on-pulse} intensity pattern.
A small section ($\sim 180 t_\mathrm{s}$; or 3 hr, if $t_\mathrm{s} = 60 s$) of this pattern
is shown in Figure 1, sampled across 1792 spectral
channels and 700 time bins (out of the simulated
duration spanning 4000 time bins).

The central spectral region, of 16.5 MHz width, is treated
as the {\it observing} band, and the associated scintillation
pattern is assumed to directly simulate {\it an observed
{\it on-pulse} dynamic spectrum}.
As an example, Figure 2 presents a zoomed portion over a short
duration ($\sim 25 t_\mathrm{s}$, or say, 25 minutes), 
where the scintle scales in both
the dimensions are clearly discernible.

Dynamic spectrum corresponding to the {\it off-pulse} region,
on the other hand,
is {\it constructed} by appropriately superposing the intensity
variation simulated across the seven bands,
following different assumed levels of genuine
unpulsed (off-pulse) emission, and those of aliasing from
contaminating bands, if any. 
For completeness, enabling a range of assessments,
we consider the following
three kinds of off-pulse dynamic spectra, as having
(a) only genuine unpulsed emission,
(b) genuine unpulsed emission plus contamination from aliasing,
and (c) no genuine unpulsed component, but with only aliased contributions.

In (a) and (b), the dynamic spectral contribution
as due to a genuine unpulsed emission is
readily obtained from the on-pulse dynamic spectrum,
suitably scaled by an assumed factor $\eta$.
Unless mentioned otherwise, $\eta$ is assumed to be 0.01.
We assume, for simplicity, that the aliased bands,
contributing in (b)  and (c), are
attenuated by also the same factor (i.e. 0.01 or -20 dB),
The off-pulse dynamic spectra, across the same span
(16.5 MHz, with 256 channels),
wherein any aliased contribution from other bands
is added together, with or without band-flips, as appropriate.
An off-pulse dynamic spectrum thus simulated for the case (c),
only with aliased contribution from only two bands adjacent
to the observed band (i.e. immediate upper and lower band)
on either side, is shown in Figure 3 (depicting a similarly
zoomed section as in Figure 2).

The Figure 4 presents the auto-correlations,
and the cross-correlation maps computed using the
the respective dynamic spectra
shown in Figure 2 and Figure 3.

The results in Table 1 are for the following
two distinct cases of simulated off-pulse dynamic spectrum;
namely, for $\eta_\mathrm{true}$ = 0.01 and  0.
In each case, the aliasing-induced contamination
from adjacent spectral bands is explicitly included
(with chosen attenuation),
for aliasing-order (AO) ranging for 1 to 3, making a total of
six versions of simulated off-pulse spectra.
These are separately used along with a common on-pulse
dynamic spectrum to estimate $\eta$ in each case.
For example, an aliasing-order ``k" corresponds to
a spectral span of k.$\Delta \nu_\mathrm{BW}$ on either side of
the observing band as being the source of contamination.

The estimates of uncertainty in $\eta$ depend of the
$N_\mathrm{eff}$, which is computed following the same procedure
as described in Section 2. In each case, having dynamic spectra
that are same for the on-pulse, but differently constructed for
the off-pulse region, the $N_\mathrm{eff}$ is computed based on the
decorrelation scales seen in the latter (i.e. off-pulse) spectra.
The decorrelation scales as seen to effectively broaden with the
number of independent spectral patterns contributing to the
constructed off-pulse dynamic spectra, as would be expected.
The simulated intensities in the dynamic spectra are essentially
exponentially distributed. These distributions would approach
to Gaussian, when additive measurement uncertainties are significant.
The overall positive bias in the estimates of $\eta$ computed from the
entire span is understood as due to the slow modulation of pulse intensity
(owing to refractive scintillations) which is shared by the
contaminants of the off-pulse dynamic spectrum.
When the suggestion made in an earlier section is followed, i.e.
$\eta$ is estimated separately for each of the shorter spans,
and such estimates combined appropriately, the average $\eta$
estimate is largely free of such bias, without loss of sensitivity.
This can be appreciated from the comparison of the
results presented in Table 1.\footnote{
The 6 models correspond to 2 sets, with and without genuine 
unpulsed emission component, 
in each of the three aliasing-orders. Two estimates of $\eta$ are presented for
each model; A: using the entire span together, and B: using
eight sub-spans separately for $\eta$ estimation, and the weighted
average of such estimates computed. The latter is largely free of the bias
corrupting the former estimates. See the main text for details.}
\begin{deluxetable}{rrrrrr}
\tablecolumns{6}
\tablewidth{0pc}
\tablecaption{
The estimated $\eta$, 
for the 6 models (2 cases each).}
\tablehead{
\colhead{AO} & \colhead{$N_\mathrm{eff}$} & \colhead{$\eta\pm\sigma_\eta$} & \colhead{$N_\mathrm{eff}$} & \colhead{$\eta\pm\sigma_\eta$}\\
   &           & $(\eta_\mathrm{true}=0.01)$ &           & $(\eta_\mathrm{true}=0)$}
\startdata
 1 A & 628       & $0.0099\pm0.0007$    &  736      & $ 0.0014\pm 0.0005$\\
 1 B &           & $0.0111\pm0.0006$    &           & $0.0011\pm0.0005$ \\
 2 A & 609     & $0.0139\pm 0.0009$   &  609      & $ 0.0043\pm  0.0008 $\\
 2 B &         &  $0.0106 \pm0.0008$  &           & $0.0007 \pm 0.0007$\\
 3 A & 517     &  $0.0137\pm 0.0012$  & 511       & $ 0.0049\pm 0.001$\\
 3 B &         & $0.0104 \pm  0.001$  &           & $0.0009\pm  0.0009$\\
\enddata
\end{deluxetable}

The $\eta$ estimates in all considered cases are consistent
with their respective model/assumed values $\eta_\mathrm{true}$
within the mentioned uncertainty. The $N_\mathrm{eff}$ changes
systematically with aliasing-order, indicating possible
increase in the decorrelation scales ($\nu_\mathrm{d}$ and $t_\mathrm{s}$).

The correlation maps in Figure 5 are presented to
illustrate the implication of the relative location of
the region corresponding to the intrinsic unpulsed/off-pulse emission,
for a location offset of $0.1R_{LC}$.
These are a result of our extended simulations, to directly obtain
special versions of off-pulse dynamic spectra
(see the text at the end of Appendix B),
and enable us to examine modification
of cross-correlation signature for different magnitudes and
specific orientations ($||$, $\perp$ and 45$^o$ to $V$)
of location offsets (for unpulsed emission source) within
the light cylinder. The corresponding correlation maps 
(such as in Figure 5), 
for different magnitude and orientation of the location offset,
indeed show the expected qualitative correspondence 
(in terms of shift and/or reduction of the correlation peak, 
as discussed in Section 2.1) in all
of the specific cases we simulated.

The above tests with simulated data demonstrate the
sensitivity of our technique in reliably searching/estimating
possible intrinsic unpulsed emission using pulsar dynamic spectra,
and confirm its desirable immunity to possible contaminants of
off-pulse dynamic spectra.

\begin{figure*}
\centering
\includegraphics[scale=0.5]{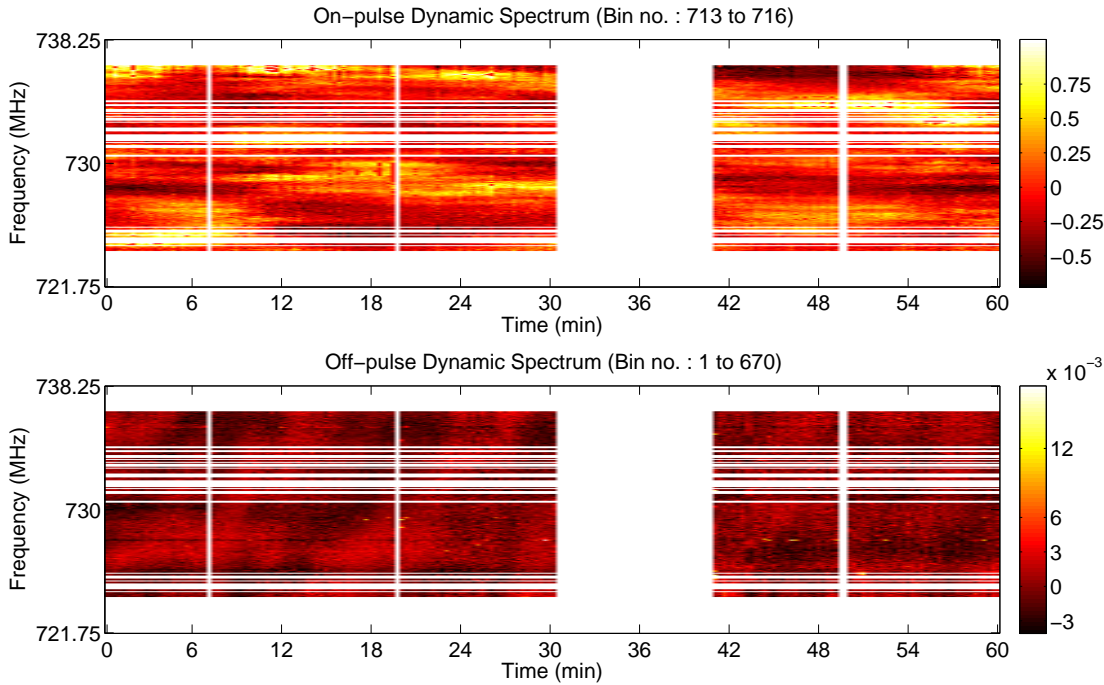}
\caption{The observed on- and off-pulse dynamic spectra
for B0329+54 at central frequency 730 MHz and bandwidth 16.5 MHz.
The horizontal
and vertical white stripes, corresponding to RFI channels and
bad time-sections, are excluded from analysis.}
\label{fig:dynspec730}
\end{figure*}

\begin{figure*}[htp]
\centering
\includegraphics[scale=0.45]{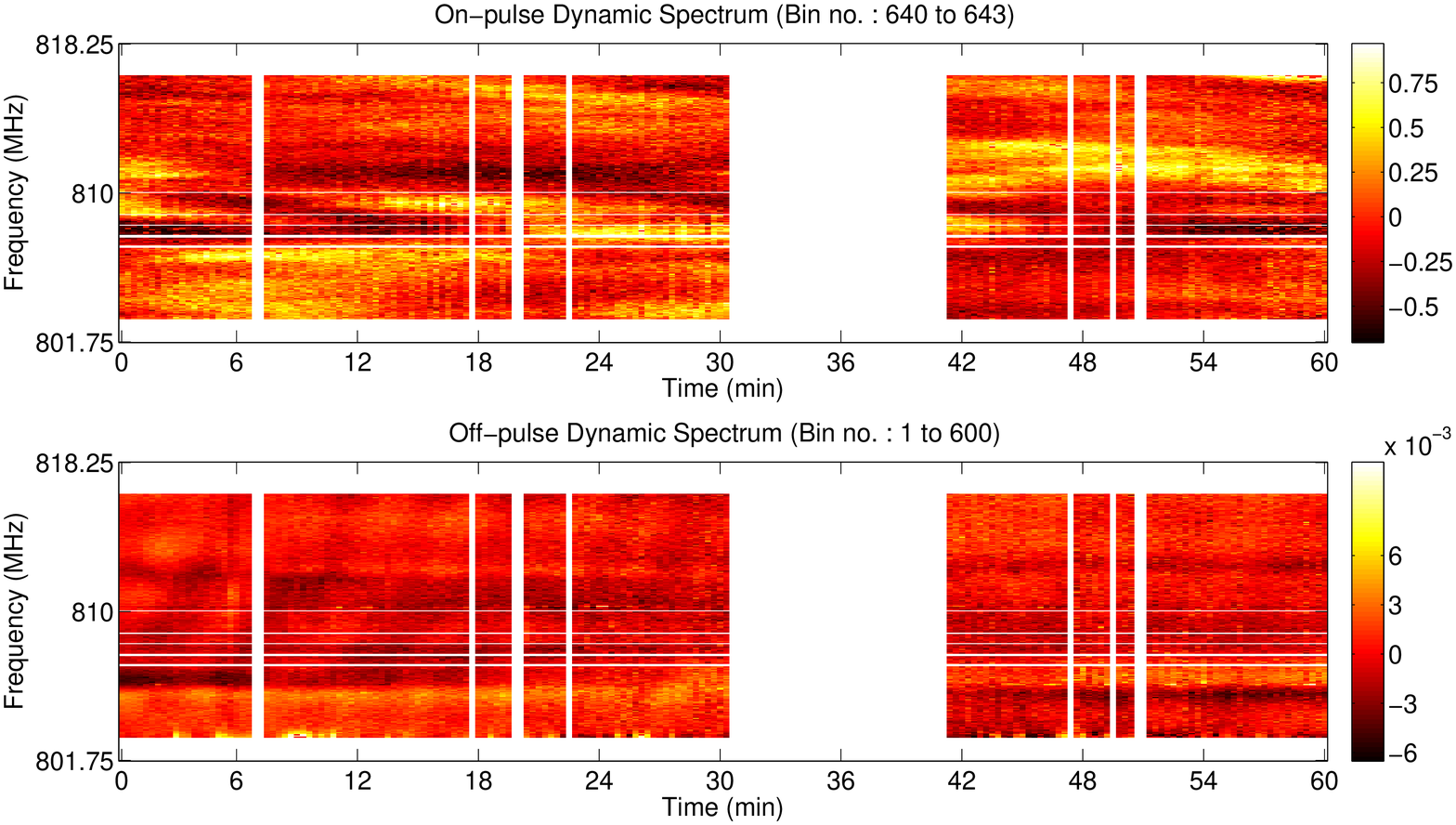}
\caption{Similar dynamic spectra as in Figure 1,
but now for the band around 810 MHz. }\label{fig:dynspec810}
\end{figure*}

\section{Illustration of our technique: A case study with data on B0329+54}

We now apply our technique to data from 
observation on pulsar B0329+54,
made using multi-band receiver system (MBR; Mann et al. 2013)
with the Robert C. Byrd Green Bank Telescope (GBT),
on July 25, 2009.

From among many pulsars observed in ten
well separated bands simultaneously,
we have chosen B0329+54 based on sensitivity considerations,
given that it is one of the brightest pulsars known.
Width of each band $\Delta\nu_\mathrm{BW}$ is 16.5 MHz,
and time-span of the observation $\Delta t_\mathrm{obs}$ is 1 hour.
Other considerations include choice of the frequency band,
as well as the resolutions in time and frequency, with which
we can expect to see scintillation features in the
the on-pulse dynamic spectrum.
These choices depend largely on the decorrelation bandwidth
and timescale, which are given by
$\nu_\mathrm{d}(\text{kHz})=59C_\mathrm{-4}^{-1.2}\lambda^{-4.4}D^{-2.2}$
and $t_\mathrm{s}(s)=149C_\mathrm{-4}^{-0.6}\lambda^{-1.2}D^{-0.6}v_7^{-1}$,
respectively (Romani et al 1985\footnote{Romani et al. (1986)
has some typographical errors in these expressions;
a wrong exponent of $C_\mathrm{-4}$ in the expression for $\nu_\mathrm{d}$, and
$v$ instead of $v_7$ the expression for $t_\mathrm{s}$(s). }).
Adequately fine sampling of the scintles, and ensuring
as large a number of scintles as possible within the spectro-temporal
span, require that $\delta\nu_\mathrm{res}\ll\nu_\mathrm{d}\ll\Delta\nu_\mathrm{BW}$
and $\delta t_\mathrm{res}\ll t_\mathrm{s}\ll\Delta t_\mathrm{obs}$.
Based on these criteria, the data at 730 MHz and 810 MHz are found suitable,
while other data at lower and higher frequencies did not meet these criteria.
For B0329+54, at 730 MHz, the estimated $\nu_\mathrm{d}$ and $t_\mathrm{s}$
are $\sim100$ kHz to $\sim1800$ kHz and 970 s to 240 s, respectively,
at 810 MHz, the corresponding $\nu_\mathrm{d}$ and $t_\mathrm{s}$ are
$\sim200$ kHz to $\sim2900$ kHz and 1100 s to 270 s,
respectively,  for $C_\mathrm{N}^2 \sim3\times10^{-5}$ and $\sim3\times10^{-4} \ m^{-20/3}$.
Available measurements by Stinebring et al. (1996) and
Wang et al. (2008) at 610 and 1540 MHz, respectively, imply
$\nu_\mathrm{d}$ and $t_\mathrm{s}$ to be about 750 kHz and 450 s, respectively,
at our lower frequency.
For comparison, our estimated decorrelation scales
(from correlation analysis such as shown in Figure 8)
of about 1 MHz, 360 s and 1.3 MHz, 400 s,
at 730 and 810 MHz, respectively, are largely consistent with the above
mentioned values, within the uncertainties.
Our dynamic spectra have frequency resolution of 64.45 kHz,
and time-resolution of 18 s, with 200 time bins across 3600 s.

The recorded raw voltage time sequences corresponding
to a bandwidth of 16.5 MHz were analyzed to obtain
dynamic spectra with frequency resolution of 64.45 kHz
(i.e., across 256 channels). After suitable corrections
for dispersion and gain compression,\footnote{
When signal levels 
are even slightly larger than the limit within which a radio receiver has linear 
response, output power becomes less than
that expected from its linear response, amounting to reduction in the gain. 
``Gain compression" refers to this reduction in gain, or non-linear response.
In the context of dispersed pulsar signals, if not corrected, such a situation 
can cause a dip proportional to the pulse intensity, with a spread in longitude
corresponding to the dispersion delay across the observed bandwidth.
This effect can contaminate off-pulse region significantly, in cases 
of bright and high dispersion measure pulsars.}
if any,
dynamic spectra for various choices
of ranges in the longitude were obtained separately.
Figures 6 and 7 show these pairs of dynamic spectra, in which
the spectral or time ranges affected by radio frequency interference
have been removed.

\begin{figure}
  \centering
  \begin{tabular}[b]{@{}p{0.45\textwidth}@{}}
    \centering\includegraphics[width=0.9\linewidth]{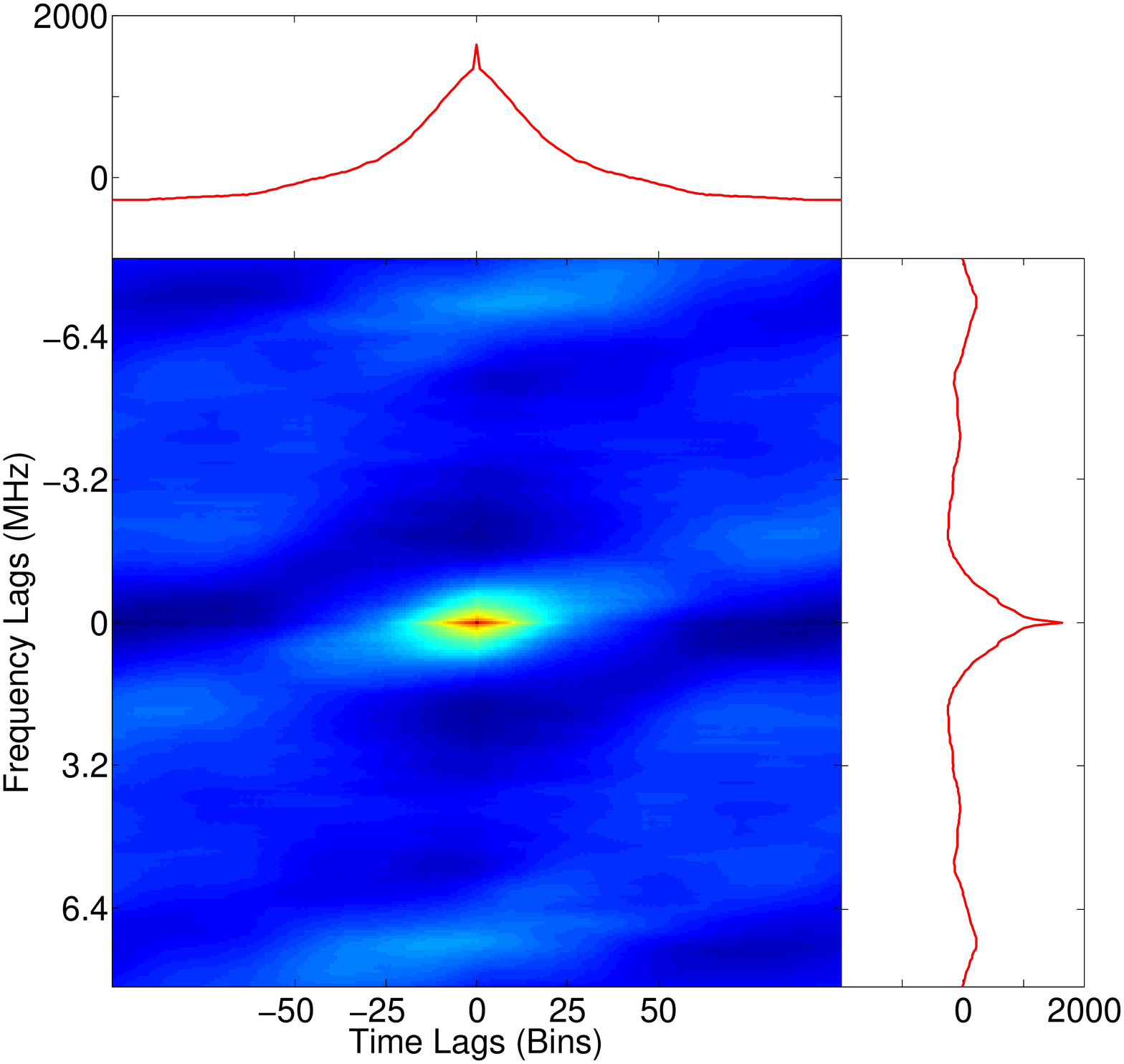} \\
    \centering\includegraphics[width=0.9\linewidth]{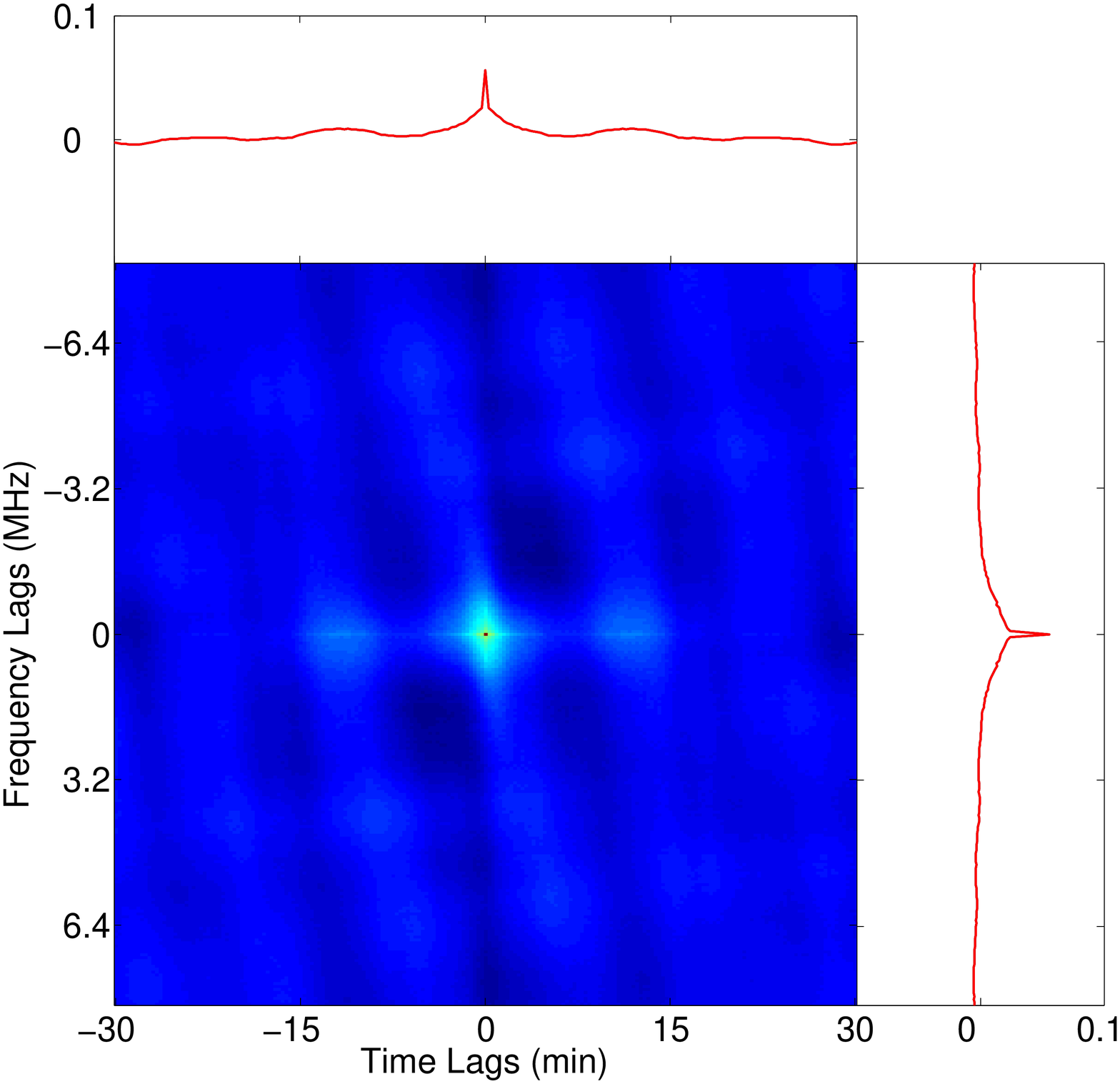} \\
    \centering\includegraphics[width=0.9\linewidth]{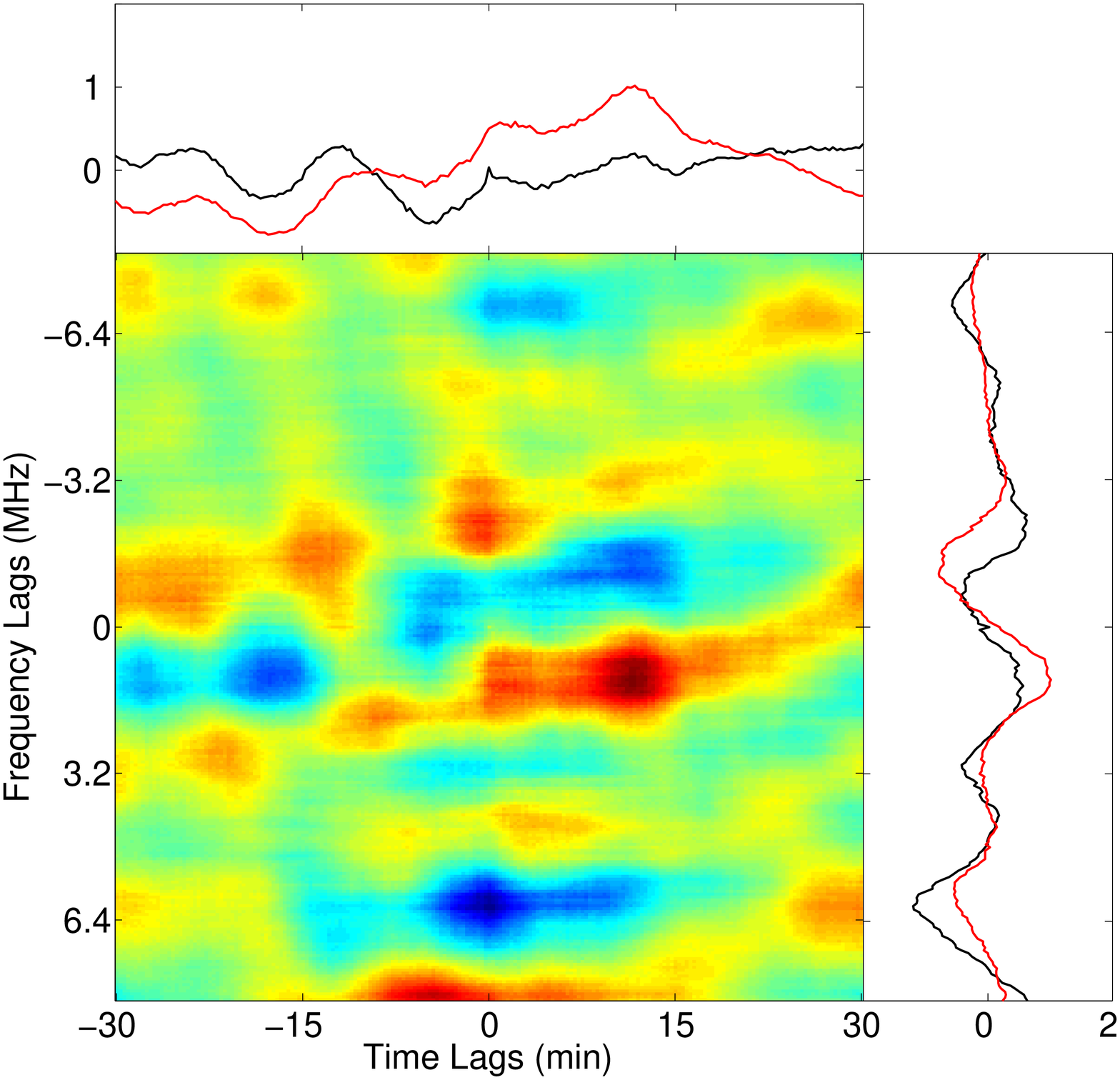} \\
  \end{tabular}
\caption{The correlation maps corresponding dynamic spectra of Figure 6. \emph{Top:} Auto-correlation of on-pulse dynamic spectra. \emph{Middle:} Auto-correlation of off-pulse dynamic spectra. \emph{Bottom:} Cross-correlation of on-pulse and off-pulse dynamic spectra. Other details of the figure are same as mentioned in Figure 4.}
\label{corr_data}
\end{figure}

For our data,
after removing bad time-sections and RFI channels,
we have performed the dynamic spectral correlations,
to estimate various quantities mentioned above, including
the measure of correlation $\eta$. More specifically,
we estimate $N_\mathrm{t},\ N_\nu,\ N_\mathrm{eff},\ \eta$ and $\sigma_\mathrm{\eta}$,
along with ${\eta}_\mathrm{max}$, as listed in Table 2.
\vspace{0.5cm}
\begin{deluxetable}{rrr}
\tablecolumns{3}
\tablewidth{0pc}
\tablecaption{Results from the dynamic spectral cross-correlation analysis of
scintillation data for the on- and off-pulse regions. The estimates
of $N_\mathrm{t}$ and $N_\nu$ assume a conservative value of the filling factor $\kappa=0.2$.}
\tablehead{
\colhead{$\nu$} & \colhead{730 MHz}   & \colhead{810 MHz}}
\startdata
$N_\mathrm{t}$ & $\sim 5 $ & $\sim 4 $\\
$N_\nu$ & $\sim 3 $ & $\sim 3 $\\
$N_\mathrm{eff}$ &  14.4 &  12.7\\
$\sigma_\mathrm{\eta}$ & 0.0017 & 0.0021\\
$\eta \pm\sigma_\mathrm{\eta}$ & 0.0002 $\pm$ 0.0017 & 0.0013 $\pm$ 0.0021\\
${\eta}_\mathrm{max}$ & 0.0045 & 0.0066
\enddata
\end{deluxetable}

As is apparent from these estimates, the off-pulse emission
can be said to be less than about 0.5\% of the main pulse
flux density (corresponding a 3-$\sigma$ limit).
The present reduction in the uncertainties in $\eta$ is
moderate, in comparison with the hard limit $\eta_\mathrm{max}$,
and is consistent with the relatively small statistics (i.e. $N_\mathrm{eff}$).

At this stage, our presentation of these results is mainly as an
illustration, but with improved spectro-temporal span of the data
(i.e. large  $N_\mathrm{eff}$),
we can expect a significant refinement in these estimations and
their uncertainties.

As can be seen readily from the
 1-d plots corresponding to the auto-correlations shown in Figure 8,
the random measurement noise de-correlates at non-zero lags,
and the sharp drop in the auto-correlation with respect to its
value at zero-lag provides a ready estimate of its relative contribution
(i.e. $\sigma^2_\mathrm{on}$ or $\sigma^2_\mathrm{off}$),
which is to be discounted
while estimating decorrelation bandwidth or time-scale.

From the cross-correlation map (the bottom plot in Figure 8), and in
general, the level of cross-correlation, or the
lack of it, can be assessed across the respective lags, and
interpreted either in terms of upper limits on the unpulsed
emission intensity, or possible separation of the associated
emission region from that of the main-pulse emission.

\section{Discussion and Conclusions}
The new technique proposed here, for searching for 
intrinsic off-pulse/unpulsed radiation for pulsars, is
inspired by the expectation that such emission
originating from  apparent location(s) matching to 
or in the {\it vicinity} of that of the pulsed component
(compared at their respective retarded emission times)
would also carry a scintillation imprint
similar to that measurable for the pulse intensity.

Needless to say that a systematic search for unpulsed emission
at a range of frequencies, with appropriate spectro-temporal
resolution  and spans, will naturally be rewarding. On the other hand,
for data sets at sufficiently nearby frequencies, a combined
estimate (or upper limit) for $\eta$ can be obtained.
For example, such an upper limit (3$\sigma$) for unpulsed
intensity of B0329+54 would be 0.4\%, based on the data at
the two frequencies.

Since our method, although a truly high angular resolution probe,
is based on longitude-resolved dynamic spectral information,
it can be expected to be applied to most of the archived
observations on pulsars, made from
single-dish and synthesis telescopes, in addition to future
observations, as long as the scintles, or the refractive
fringing when present, are at least Nyquist sampled.
In fact, it would not be surprising to see a massive initiative
to search of the continuous/unpulsed emission in the near future,
using existing data and new observations.

Although the relevant correlation would reduce exponentially 
with increasing apparent angular offset $\Delta S/z$ (Cordes et al. 1983) 
of the source to be searched, opportunity to detect significant peak
in the correlation map (at a non-zero lag in time) is to be expected
when the orientation of the offset is along the pattern velocity
(or within an alignment margin of $(S_\mathrm{d}/\Delta S)^c$),
as illustrated in our simulations.
Interestingly, given the reported pulsar spin-velocity alignment,
for young pulsars in particular (Johnston et al. 2005),
the location offset along latitudinal direction is likely to
be along the pattern velocity, when the scintillation speed
is dominated by pulsar motion.
Regardless of the possible apparent location of the region
responsible for unpulsed emission, if any, within
the light-cylinder, it is unlikely that the associated
key morphology (say, w.r.t. rotation axis) would
differ significantly from pulsar to pulsar.
This combined with the expected variety in
the orientation of the {\it apparent}
velocity of the diffractive scintillation pattern
(again viewed with respect to the pulsar spin-axis) and
in $S_\mathrm{d}$ (depending on sight-line and frequency),
suggests that there would be adequate number of
known pulsars offering conducive situation
for the proposed probe to be rewarding, in either
detecting their elusive continuous emission, or
ruling it out to the extent possible.

In summary, the existence of unpulsed emission from
pulsars is yet to be fully established, let alone be understood.
However, if detected unambiguously, any unpulsed radio radiation
intrinsic to pulsars would indeed be
a precious token of the mysteries of
emission mechanism in radio pulsars that are
yet to be unraveled,
and could provide important missing clue
to further our understanding of the key processes at work.
Our proposed detection method, exploiting
the resolving power of the interstellar telescope,
as a powerful tool for reliable and sensitive search/detection
of unpulsed/off-pulse emission, should open a new window for this
promising exploration.

{\acknowledgements We gratefully acknowledge contribution from 
Karishma Bansal in the very early phase of this work 
(during her Visiting Studentship at the Raman Research Institute). 
We thank our anonymous referee for constructive comments 
and suggestions.}


\appendix


\section{A: Aliasing in dynamic spectra,
and possible contamination in the off-pulse region}

In this section, we discuss and illustrate how the observed data,
and the off-pulse region in particular,
be affected if the spectral filtering in the
receiver chain were to be imperfect, allowing a non-zero fraction
of sky signal out side the band of interest to pass through, 
even though attenuated significantly. Although, in practice,
a variety of non-idealities in spectral filtering are possible\footnote{
The key filtering stages include, a) image-band rejection before heterodyne 
or mixing stage, and b) band-defining before digitization (at Nyquist rate)
of signal either at baseband or that located around a chosen center 
frequency (where harmonic or band-pass sampling at Nyquist rate is employed).},
for illustration purpose, we consider a simple case of
the band-defining filter having a non-zero response
in adjacent spectral ranges on either side of the intended
band of observation. 

It is easy to see that the contribution
from a dispersed pulse in these out-of-band spectral regions
would be aliased in the Nyquist sampled band of interest. On dedispersion,
these aliased contributions from the {\it folded}  bands
would not only spill out side the pulse window,
but can systematically spread across
a large part of the off-pulse longitudes,
and could disguise as off-pulse emission.
For an illustrative example of this effect,
lets assume that the filter function over
the desired band (of width = $BW$) is flat,
as corresponding to a perfect rectangular filter,
but this non-ideal filter offers finite, though high,
attenuation in other spectral regions.  In our simulation,
we consider the out-of-band region on
either side to be 3 times wider than $BW$,
and relative attenuation to be 20 dB,
such that the filter response $F(\nu)$
as function of frequency $\nu$ is given by
\begin{equation}
F(\nu)=\left\{
\begin{array}{lc}
1 \,,& |\nu-\nu_\mathrm{c}|\leq 0.5BW\,,\\[3pt]
0.01 \,,& 0.5BW\leq|\nu-\nu_\mathrm{c}|\leq 3.5BW\,,\\[3pt]
0 \,,& otherwise\,\\[3pt]
\end{array}
\right.
\end{equation}
where, $\nu_\mathrm{c}$ is the center frequency of observation.

\begin{figure}
\centering
\includegraphics[scale=0.40]{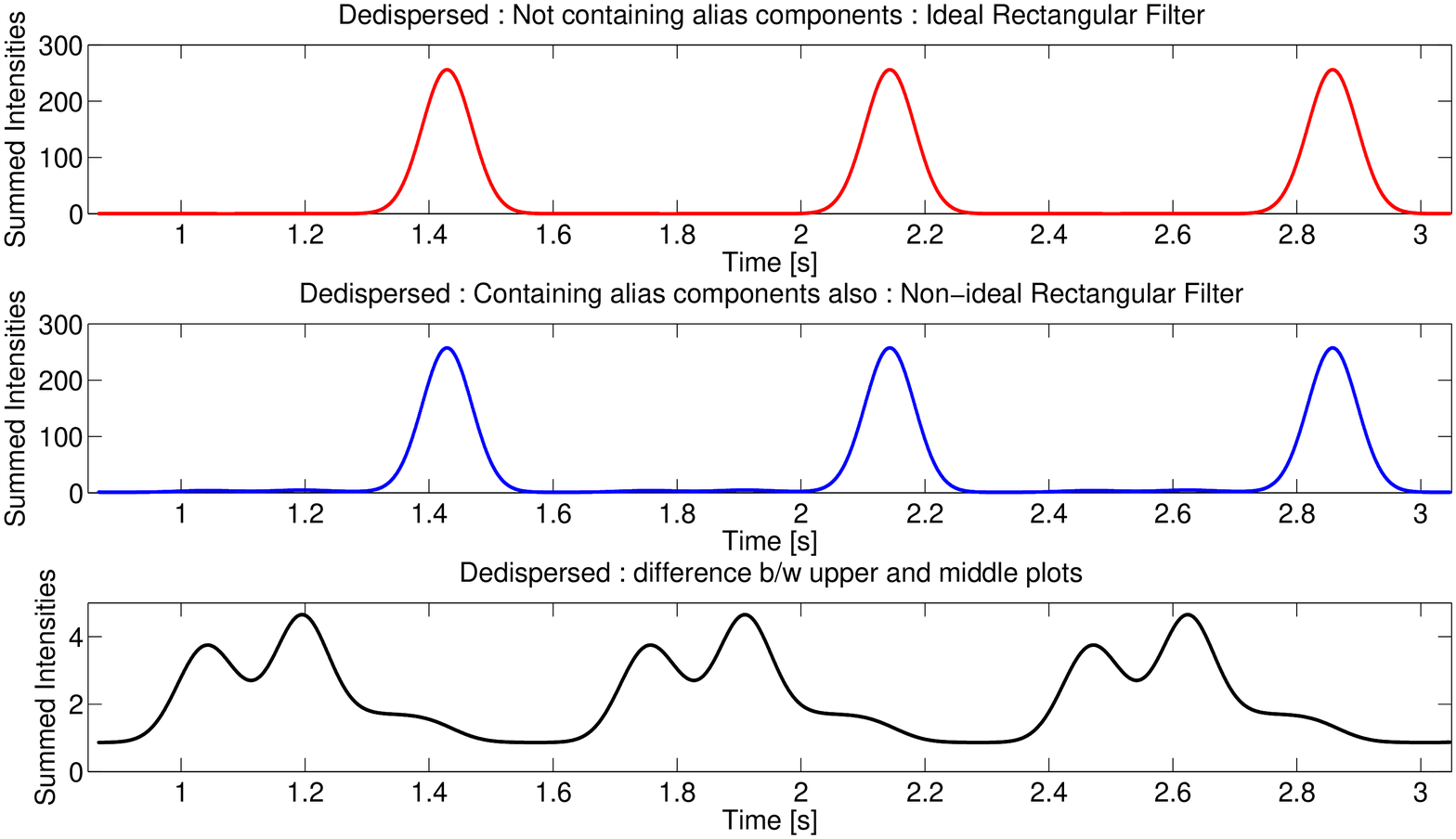}
\caption{A simulated dedispersed pulse sequence in
the case of an ideal filter function is shown in the top panel,
for reference. A similarly simulated and dedispersed
time sequence resulting from aliasing from adjacent bands
is presented in the middle panel. The difference between
these two sequences (as shown in the bottom panel)
provides an illustration of the contamination due to
aliasing effects. For details,
please refer to the text in this appendix.
}
\label{}
\end{figure}

Assuming a train of Gaussian pulses from a pulsar,
the intensity pattern across time and frequency
can be expressed as
\begin{equation}
I(\nu,t)=I_0\exp\{-((t-\tau(\nu))-nP)^2/(2\sigma^2)\}
\end{equation}
where, $P$ is the pulsar period,
$\sigma$ is the standard width of the Gaussian,
$n$ is number of periods defining the time span
of the simulated sequence. The dispersion delay at
frequency $\nu$, with respect to that at a
reference frequency $\nu_\mathrm{ref}$, is given by
\begin{equation}
\tau(\nu)=4.15\times10^3DM\left[\frac{1}{\nu^2}-\frac{1}{\nu_\mathrm{ref}^2}\right]\ (s)
\end{equation}
where $DM$ is the dispersion measure in
$pc\ cm^{-3}$ and the frequencies are in MHz.
A dynamic spectrum containing dispersed pulses
was simulated assuming $BW$= 16 MHz centered at
$\nu_\mathrm{c}$= 300 MHz, $\sigma$ = 0.04 s, and
for pulsar parameters similar to that of B0329+54
($DM$ = 26.77 $pc\ cm^{-3}$, $P$= 0.714472578 s).
\par
The top panel of Figure A1 (or 8) shows the dedispersed
pulse sequence when no aliasing of sky signal occurs from outside
of the desired band of bandwidth $BW$ (a case of ideal filtering).
The middle panel shows a similarly obtained sequence,
but now including aliasing (of remaining small level of sky signal) 
from the adjacent bands
as described above. Note that the dispersed pulse contribution
from the aliased adjacent bands would be at a much lower level,
but will make its appearance in the off-pulse region.
These contributions will occupy different longitude
spreads after dedispersion,
depending on the alias order,
in addition to the $DM$ and $P$.
Significant contamination in both the on-pulse and
the off-pulse regions, as result of aliasing,
is apparent from the difference between the sequences
in the top two panels, as shown in the bottom panel.
Here, we have deliberately used a lower $\nu_\mathrm{c}$ (=300 MHz),
than those for the data we present (i.e. 730 MHz or 810 MHz),
so that the mentioned contamination is more pronounced.

\section{B: Simulation of Diffractive Interstellar Scintillation (DISS) and Dynamic Spectra}
The two main steps in simulation of DISS dynamic spectra,
using a thin-screen approximation, are:
(i) generation of a random phase screen following an assumed
spatial distribution of electron density irregularities in the intervening
ISM, which modifies the emerging wave-front, and
(ii) calculation of resultant intensity, of the received signal
at the observer's location, as a function frequency and time.

\par As mentioned already, the most accepted 3D spatial power spectral description
of the turbulent ISM is a power-law spectrum across spatial
frequency $q$, ranging from $q_\mathrm{min}$ to $q_\mathrm{max}$, (Armstrong et al. 1995)

\begin{equation}
P_\mathrm{3N}(q_x,q_y,q_z)=C_\mathrm{N}^2\,\, q^{-\beta};\ q=\sqrt{q_x^2+q_y^2+q_z^2}
\end{equation}
where, $C_\mathrm{N}^2$ is the level of turbulence, and the power-law index
$\beta=11/3$ for the Kolmogorov turbulence. Armstrong et al. (1995)
have given evidence of the validity of Equation B1 for
$q_\mathrm{min}=10^{-12}\ m^{-1}\ <\ q\ <\ q_\mathrm{max} = 10^{-6}\ m^{-1}$,
and have derived the typical turbulence strength
$C_\mathrm{N}^2\sim10^{-3}\ m^{-20/3}$, but the $C_\mathrm{N}^2$ can deviate
significantly from this typical value, depending on direction
and distance to the source.
\par A convenient way to study propagation effects due to this
3-D distribution of refractive index irregularities in the ISM
is to model the modification of the incident wave-front by an
equivalent thin phase-changing screen, located between the
source and the observer. 
We use this thin screen approximation (see Lovelace 1970; Romani et al. 1986)
for our present
simulations, wherein the equivalent 2-D spatial power spectrum ($P_\mathrm{2\phi}$)
of the phase deviation $\phi$ is given by
\begin{eqnarray}\nonumber
P_\mathrm{2\phi}(q_x,q_y)&=& 2\pi z(\lambda r_\mathrm{e})^2\,\, P_\mathrm{3N}(q_x,q_y,q_z=0)\\
&=& 2\pi z(\lambda r_\mathrm{e})^2C_\mathrm{N}^2\left(\sqrt{q_x^2+q_y^2}\right)^{-\beta}
\end{eqnarray}
where, $z$ is the distance to the source,
$r_\mathrm{e}=2.82\times10^{-15}$ m is the classical radius of the electron,
and $\lambda$ is the wavelength of the propagating radiation.

\subsection{Generation of Equivalent Thin Phase-Screen using FFT-Based Technique}
The spatial power spectrum of the equivalent 2-D thin screen
and the associated distribution of the random phase deviation
$\phi(x,y)$ across that screen in transverse plane $(x,y)$
have the following Fourier relationship
\begin{equation}
\phi(x,y)=\int_\mathrm{-\infty}^{+\infty}\int_\mathrm{-\infty}^{+\infty}g(q_x,q_y)\sqrt{P_\mathrm{2\phi}(q_x,q_y)}\exp\left[-j2\pi(xq_x+yq_y)\right]dq_xdq_y
\end{equation}
i.e., $\phi(x,y)$ is the (inverse) Fourier transform of
the product $g(q_x,q_y)\sqrt{P_\mathrm{2\phi}(q_x,q_y)}$,
where $g(q_x,q_y)$ is a Hermitian-symmetric complex
Gaussian variable representing zero-mean white noise process,
with unity variance (Johansson \& Gavel 1994).
We obtain $\phi(x,y)$ distribution using the above relation,
employing FFT technique for computational ease.

The discrete form needed for simulation,
of the Equation B3, for a square screen, $\phi$,
made-up of $N\times N$ grid points, is
\begin{equation}
\phi=2\pi(2\pi)^{-\beta/2}(N\Delta r)^{-1+\beta/2}\sqrt{2\pi z(\lambda r_\mathrm{e})^2C_\mathrm{N}^2}\left[\mathcal{FT}^{-1}\left\{g\,M_0\right\}\right]
\end{equation}
where, $\Delta r$ is spatial sampling interval,
$g$ is a $N\times N$ matrix (the procedure to obtain
it is explained in the following) and $M_0$ is also
a $N\times N$ matrix whose elements are
$M_0(i,j)=[(i-N_\mathrm{c})^2+(j-N_\mathrm{c})^2]^{-\beta/4}$,
where the origin is defined at $N_\mathrm{c} = N/2 + 1$,
and the contribution at zero spatial frequency
is set to zero, i.e. $M(N_\mathrm{c},N_\mathrm{c})=0$.
The recipe for getting the
Gaussian random matrix $g$ is as follows:
\begin{enumerate}[(i)]
\item generate a complex matrix (say) $M$ of size $N\times N$,
whose elements are $a+j\,b$, where, $j=\sqrt{-1}$, $a$ and $b$
are independent Gaussian random numbers with zero mean and unity variance.
\item obtain the 2-D discrete Fourier transform of `$M$',
which will be the required matrix $g$ ($g=\mathcal{FT}^{-1}\left\{M\right\}$).
\end{enumerate}
\subsection{Electric Field Distributions and Resultant Intensity in the Observer's Plane}
Having generated $\phi(x,y)$ (i.e., the discrete $\phi$, Equation B4),
the electric field distribution at the thin screen can be given by
\begin{equation}
E_\mathrm{s}(x,y)=\exp(-j\,\phi(x,y)).
\end{equation}
In the thin screen approximation, the ray optics is applicable
and so the electric field received at any point $(\xi,\eta)$ on
the observer plane, can be represented by the Fresnel-Kirchhoff
integral (Born \& Wolf 1980)
\begin{equation}
E_\mathrm{O}(\xi,\eta)= \frac{e^{-j\pi/2}e^{j2\pi z/\lambda}}{2\pi r_\mathrm{f}^2}\int\int \exp\left[j\phi(x,y) +j\frac{(x-\xi)^2+(y-\eta)^2}{2r_\mathrm{f}^2}\right]dxdy
\end{equation}
where, $r_\mathrm{f}=\sqrt{\lambda z/2\pi}$. This integral can be either
calculated from 2-D numerical integration or methods using
Fourier transform. We have used angular spectrum method to
calculate the electric field, i.e., via following relation (in discrete form)
\begin{equation}
E_\mathrm{O}=\mathcal{IFT}\{\mathcal{FT}\{h(x,y)\}\,\,\,\mathcal{FT}\{\exp(-j\phi)\}\}
\end{equation}
where, $h(x,y)$ is called transfer function. By use of
the above method to get $E_\mathrm{O}$, the spatial sampling interval
at the observer's plane will be the same $\Delta r$ as
that of phase distribution of the thin phase screen.
So corresponding to each value of input frequency/wavelength,
we will have phase distribution $\phi$ of the thin screen
(in discrete form a matrix of size say $N\times N$)
and the electric field $E_\mathrm{O}$ at the observer's plane
(again a matrix of size $N\times N$). From this matrix $E_\mathrm{O}$,
we select a spatial 1D cut, say $E_\mathrm{O}(r)$, which may be an
arbitrarily chosen row or column,
and obtain $E_\mathrm{O}(\nu,r)$ by varying only $\nu$
in uniform steps over the range of interest.
The spatio-spectral description of observed intensity is trivially
obtained as $I_\mathrm{O}(\nu,r)$ = $|E_\mathrm{O}(\nu,r)|^2$.
This $I_\mathrm{O}(\nu,r)$ can be
translated into $I_\mathrm{O}(\nu,t)$, i.e. the dynamic spectrum,
by assuming a velocity $V_\mathrm{trans}$ along $r$ for the intensity pattern
in the observer's plane, which depends on the relative
transverse velocities of the pulsar, the scattering medium and the observer.

\subsubsection{Our Simulation Parameters and Results}
Diffractive effects correspond to spatial frequency range
$q\sim10^{-8}\ m^{-1}$ to $10^{-6}\ m^{-1}$ of the ISM
irregularities
(Stinebring 1996; Rickett 1988; Narayan 1988; Wang, Manchester 2008 ).
We have used a square (scattering) screen so sampling interval, say $\Delta r$,
in x and y directions are equal, i.e. $\Delta x=\Delta y =\Delta r$.
The Nyquist sampling criteria demands $\Delta r=1/(2q_\mathrm{max})$.
What size of the phase screen will be suffice for simulation of DISS ?
The observer receives radiation from a cone of half-angle
$\theta_\mathrm{S}\approx r_\mathrm{ref}/z\approx r_\mathrm{mp}/z\approx \lambda/(2\pi S_\mathrm{d})$,
(Cordes 1986) where, $r_\mathrm{ref}$ is refractive length scale, $r_\mathrm{mp}$
is multi-path propagation length scale (strong scintillation),
$S_\mathrm{d}$ is diffractive length scale, $z$ is distance from the
observer to the thin screen and $\lambda$ is the wavelength of the
radiation from pulsar. So to properly simulate DISS,
the phase screen size $r_\mathrm{max}$ should be at least
$\sim r_\mathrm{mp}\approx r_\mathrm{f}^2/S_\mathrm{d}$ in each of the two dimensions.
Hence the required number of grid points $N$ across $r_\mathrm{mp}$,
for a $N\times N$ matrix describing the screen,
would be $N \geq r_\mathrm{max}/\Delta r = 2q_\mathrm{max}r_\mathrm{f}^2/S_\mathrm{d}$
Thus for the case of our data on B0329+54, where
$\nu$=810 MHz and  $z=1.44$ kpc,
$r_\mathrm{mp} \sim 10^{11} $ m.
To satisfy the criteria $\Delta r\leq 1/(2q_\mathrm{max})$,
the required $N\sim2^{18}$ [$N=10^{11}/(5\times10^5)\sim 2\times10^5 \sim2^{18}$] !
To generate a phase screen of this overwhelmingly large dimension,
of order $2^{18}\times 2^{18}$, and the subsequent Fourier analysis
involving bigger dimension (i.e., $2^{19}\times2^{19}$)
is not only computationally intensive (even with use of FFTs),
but well beyond the readily available computing resources.

However, we note the {\it red} nature of the underlying spatial
spectrum of phase variation
$P_\mathrm{2\phi}\propto q^{-\beta}$ ($\beta$ is +ve).
The associated structure function for phase $\phi$, at a given scale $r$
can be expressed as $D_\mathrm{\phi}(r) = (1^c)^2 (r/S_\mathrm{d})^{(\beta -2)}$, given that for
the diffractive (or the coherence) scale $S_\mathrm{d}$,  the phase structure
function is 1 $radian^2$ (Armstrong et al. 1995).
Since contribution to phase fluctuations from smaller spatial scales
is expected to decrease rapidly, for the relevant values of $\beta$,
we consider revision of the sampling scale $\Delta r$, such that
$D_\mathrm{\phi}(\Delta r) \leq {\Delta \phi}^2_\mathrm{min}$, for the desired small
phase variation ${\Delta \phi}_\mathrm{min}$ that is to be sampled duly.
With this criterion, and recalling Equation 8,
we express the required grid dimension as

\begin{equation}
N \geq \frac{(\nu/\nu_\mathrm{d})}{({\Delta \phi}_\mathrm{min})^{2/(\beta -2)}}  
\end{equation}

For example, with ${\Delta \phi}_\mathrm{min}$ = 0.1 radian, and $\beta$ = 11/3,
$N \geq 16 (\nu/\nu_\mathrm{d})$. Choosing a suitable ratio $(\nu/\nu_\mathrm{d})$, say 200,
the requirement of $N \geq 3200$ appears feasible with computational constraints,
i.e. without needing supercomputers, and more importantly,
without compromising significantly on the details of the phase screen.

In our present simulations, we have used
$\nu_\mathrm{d}\sim 1.35$ MHz, to keep reasonable correspondence with
the discussed observations, but use a relatively smaller $\nu$ of 270 MHz.
We use $N=4096$, so that the screen and the diffraction patterns are sampled
with adequate details (corresponding to $\Delta r=5\times10^5$ m and
$r_\mathrm{max} \sim 2\times10^{9}$)\footnote{It is worth noting that now
the Fresnel scale $r_\mathrm{f}$ and $z$ are artificially small
and $C_\mathrm{N}^2$ is correspondingly
large, as a result of the spatial dynamic range we have chosen.},
and the resultant dynamic spectrum
suffices for demonstrating the key aspects of our technique.
It is worth pointing out that now
the Fresnel scale $r_\mathrm{f}$ and $z$ are artificially small and $C_\mathrm{N}^2$ is correspondingly
large, as a result of the spatial dynamic range we have chosen.
However, the scale of direct relevance to us, that is the
diffraction pattern scale $S_\mathrm{d}$ corresponds to typical 4 samples
across $r$, implying $(t_\mathrm{s}/4)$ as the sampling interval in the dynamic spectrum.
The overall time scale, and $t_\mathrm{s}$, can be defined by choice of the
velocity $V_\mathrm{trans}$, if required.
In any case, the simulated time span corresponds to $\sim 1000 t_\mathrm{s}$.

For completeness, to examine the sensitivity of the
correlation technique to the relative location off-pulse emission region,
we extended our simulations to obtain the off-pulse dynamic spectra
separately from that for the pulsed component.
Using the common description for the phase pattern,
we added a suitable extra phase-gradient to it for
simulating new phase screen, corresponding to the location offset,
and used the resultant intensity pattern for constructing
off-pulse dynamic spectra.
The magnitude and direction of the location offset were varied,
and the resultant cross-correlation maps were examined (see Figure 5).

\end{document}